\documentstyle[12pt,pictex,epsfig,amssymb,amsbsy]{article}

\title{{Wilson Loops %%% \\[.2cm]
and QCD/String \\Scattering Amplitudes}\\[1cm] }

\author{
Yuri Makeenko%
\footnote{Also at the Institute for Advanced Cycling,
Blegdamsvej 19, 2100 Copenhagen \O, Denmark}
\\[.2cm]
{\normalsize
{\it Institute of Theoretical and Experimental Physics}} \\
 \normalsize {\it 117218 Moscow, Russia}\\
\normalsize{{makeenko@itep.ru}}
 \\[.4cm] and \\[.4cm]
Poul Olesen$^*$ \\[.2cm]
 \normalsize{\it The Niels Bohr International Academy}\\
\normalsize{\it The Niels Bohr Institute}\\
  \normalsize {\it Blegdamsvej 17, 2100 Copenhagen \O, Denmark}\\
\normalsize{polesen@nbi.dk}}

\date{\mbox{}}

\renewcommand{\thefootnote}{\fnsymbol{footnote}}
\newcommand{\newsection}{
\setcounter{equation}{0}
\section}

\def\appendix#1{
  \addtocounter{section}{1}
  \setcounter{equation}{0}
  \renewcommand{\thesection}{\Alph{section}}
  \section*{Appendix \thesection\protect\indent \parbox[t]{11.715cm} {#1}}
  \addcontentsline{toc}{section}{Appendix \thesection\ \ \ #1}
  }

\def\e{{\,\rm e}\,}
\def\d{{\rm d}}

\def\i{{\rm i}}
\def\K{K}
\def\t{s}
\def\s{t}
\def\Dp{{\cal D}_{\rm diff}\,}
\def\eol{\hspace*{\fill}\linebreak}

\newcommand{\rf}[1]{(\ref{#1})}
\newcommand{\eq}[1]{Eq.~(\ref{#1})}
\def\be{\begin{equation}}
\def\ee{\end{equation}}
\def\beq{\begin{equation}}
\def\eeq{\end{equation}}
\def\bea{\begin{eqnarray}}
\def\eea{\end{eqnarray}}
\def\LA{\left\langle}
\def\RA{\right\rangle}
\def\D{{\cal D}}
\def\Tau{{\cal T}}
\newcommand{\non}{\nonumber \\*}
\newcommand{\ie}{{i.e.}\ }

\newcommand{\ppint}{\int\limits_0^{2\pi}\hspace{-1.2em}\not\hspace{.75em}}

\def\lesssim{\mathrel{\mathpalette\fun <}}

\def\fun#1#2{\lower3.6pt\vbox{\baselineskip0pt\lineskip.9pt
\ialign{$\mathsurround=0pt#1\hfil##\hfil$\crcr#2\crcr\sim\crcr}}}
\def\la{\lesssim}

\begin{document}

\begin{titlepage}

\maketitle

%\vspace*{-12cm}
%\begin{flushright}
%%ITEP--TH--13/09\\
%%NBI--HE--09--??\\
%%%hep-th/yymmnnn\\
%%March, 2009
%%\end{flushright}
%%\vspace{11cm}

%\vskip .5 cm
\begin{abstract}
We generalize  modern ideas about the duality between Wilson loops and 
scattering amplitudes in ${\cal N}=4$ SYM to large $N$ QCD
by deriving a general relation between QCD meson scattering amplitudes and
Wilson loops. We then investigate properties
of the open-string disk amplitude integrated over reparametrizations. 
When the Wilson loop is
approximated by the area behavior, we find that the QCD scattering amplitude
is a convolution of the standard Koba--Nielsen integrand and a kernel.
As usual poles originate from the first factor, whereas no (momentum
dependent) poles can arise from the kernel. 
We show that the kernel becomes a constant when the number of external 
particles becomes large. The usual Veneziano amplitude then emerges in
the kinematical regime where the Wilson loop can be reliably 
approximated by the area behavior. In this case we obtain a direct duality 
between Wilson loops and scattering amplitudes when spatial variables and 
momenta are interchanged, in analogy with the $\cal N$=4 SYM case.

\end{abstract}

\thispagestyle{empty}
\end{titlepage}
\setcounter{page}{2}
\renewcommand{\thefootnote}{\arabic{footnote}}
\setcounter{footnote}{0}

\newsection{Introduction}

The relation between planar diagrams and dual resonance models has
a long history since the pioneering works~\cite{NO70}. 
A long-standing belief~\cite{Hoo74}
is that $SU(N)$ Yang--Mills theory
is equivalent at large $N$ to a free string, while the $1/N$-expansion  
corresponds to interactions of the string.%
\footnote{See e.g.\ Ref.~\cite{Mak02} for an introduction and 
review of the old works on the QCD/string correspondence.} 
A great recent progress along this line 
is associated for ${\cal N}=4$ super Yang--Mills theory (SYM) with 
the AdS/CFT correspondence~\cite{Mal98a,GKP98}
(see Ref.~\cite{AGMOO99} for a review), 
where the strong-coupling limit
of SYM is described by supergravity in anti-de Sitter space 
$AdS_5 \times S^5$. 

The (finite part of the) 4-gluon Maximally-Helicity-Violating 
on-shell scattering amplitude in SYM theory has the form 
%%%%structure
\be
A(s,t)=A_{\rm tree}\,\e^{f(\lambda) \log^2 (s/t)}
\label{BDS}
\ee
(where $s$ and $t$ are usual Mandelstam's variables)  
as was conjectured~\cite{BDS05} on the basis
of three-loop calculations.
To explain \eq{BDS}, the 
Wilson-loop/scattering-amplitude (WL/SA) duality 
was introduced~\cite{AM07a} at large 't~Hooft
couplings $\lambda$, which has been then advocated in SYM perturbation
theory~\cite{DSK07}. This duality (for a review see Ref.~\cite{AR08})
states that the scattering
amplitude (divided by the kinematical factor $A_{\rm tree}$) equals
the Wilson loop for a polygon whose vertices $x_i$ are related to
the momenta $p_i$ of scattering gluons by
\be
p_i= \K\left( x_i-x_{i-1}\right),
\label{du}
\ee
where $\K=1/2\pi\alpha^\prime$ is the string tension.

Our goal in this paper is to find out what features of the described
WL/SA duality (if any) remain valid for QCD and, in particular,
how is it possible to maintain the relation of the type~\rf{du}
which would relate {\em large}\/ momenta in scattering amplitudes with
loops of {\em large}\/ size. Of course this is not possible in QCD
perturbation theory, 
where $|p|\sim 1/|x|$ because of
dimensional ground. But nonperturbatively a dimensional parameter 
$\K\approx (400~{\rm MeV})^2$
appears in QCD, which shows up
in the area-law behavior of asymptotically large Wilson loops:
\be
W(C)\stackrel{{\rm large}~C} \propto \e^{-\K S_{\rm min}(C)} \,,
\label{a-l}
\ee
where $S_{\rm min}(C)$ is the area of the minimal surface bounded by $C$,
that results in confinement. Strictly speaking, this requires 
the limit of the large number of colors 
$N$ or the quenched approximation.

As is well-known by now, 
a string theory, which QCD is supposedly equivalent
to, is not the simplest Nambu--Goto string. Some
extra degrees of freedom living on the string are required
which are most probably
conveniently described by a presence of extra dimensions. The asymptotic
behavior~\rf{a-l} is nevertheless universal for large loops. 

In this paper the main topic is an investigation of the relation between the 
Wilson loop and the corresponding amplitude in large $N$ QCD, or, 
alternatively, quenched QCD for any $N$. In general 
this involves integration over an infinite number of loops. However, if
one considers large loops there is a considerable amount of evidence from
lattice gauge calculations in 2+1 and 3+1 dimensions for various $N$'s
that the Nambu--Goto action describes the behavior of the Wilson loops
quite well. To give an example, in \cite{lw02} it is shown with
unprecedented precision that the static 
quark potential in quenched $SU(3)$ lattice gauge theory is well described by 
the first two terms ($\propto r$ and the L\"uscher term
$\propto 1/r$) in a long-distance expansion of the Nambu--Goto action.
Furthermore, in both three and four dimensions the transition
from perturbative to string behavior takes place ``at surprisingly
small distances'' \cite{lw02}. There also exists a number of
other comparisons between results from the Nambu--Goto action, e.g. 
the closed 
string spectrum and $SU(N)$  for various values of $N$, see \cite{close}
where further references can be found.

The various results from lattice gauge theories can be summarized by the
statement that the Nambu--Goto action describes the large and not so
large \cite{lw02} distance behavior of quenched QCD surprisingly well. This
action has the well known anomaly for $d\neq 26$, which however is suppressed
for long strings \cite{p85}. The remarkable success of the Nambu--Goto string
as an effective action led us to reconsider the relation between the Wilson
loop $W(C)$ and the corresponding amplitude. The idea is to perform the 
unpleasant sum over all $C$'s by inserting the Nambu--Goto action in $W(C)$.
Then the sum over $C$ becomes an integration over the string field
$x^\mu (\sigma, \tau)$, and is at least in principle controllable.
In practice, in this paper we start out with a more modest program, where
only the area behavior of $W(C)$ is inserted in the form of the disk
amplitude. Thus, the effects of the L\"uscher term is not included in the
investigation reported in this paper.

The present paper is an extended version of Ref.~\cite{MO08}, 
but some of the reported results are novel. 
In Sect.~2 we derive a relation between the $M$-meson scattering amplitude 
and the Wilson loop for the case of fermion and scalar quarks, valid for
QCD in the large $N$ limit. Our results are general in the sense that
if the Wilson loop is known one can obtain the scattering amplitudes by
performing some path integrations. In Sect.~3 we then take up the 
old idea that the large $N$ QCD Wilson loop should be identified with
the disk amplitude in certain string models as far as the leading
large distance behavior is concerned. We emphasize that it is crucial
in this construction to integrate the string disk amplitude over 
reparametrizations of the boundary contour.
In Sect.~4 we show, 
taking a functional Fourier transform to momentum space, how the disk
amplitude leads to the Koba--Nielsen amplitude. 
Integrating over the reparametrizations, we also derive
projective-invariant off-shell scattering amplitudes.

In Sect.~5 we return to the general
formula from Sect.~2 for the relation between the meson (made from
fermion quarks) scattering amplitude and the Wilson loop. Here we insert
the area law (with no subleading  perimeter term, L\"uscher term, ...)
and for the (off-shell) $M$-particle amplitude we derive the formula
\begin{eqnarray}
&&G(\Delta p_1,...,\Delta p_M)\propto 
\prod_1^{M-1}\int\limits_0^{\phi_{i+1}}\d\phi_i\, 
\prod_{j=1}^M
\left[\frac{\sin[(\phi_{j+1}-\phi_j)/2]\sin[(\phi_{j}-\phi_{j-1})/2]}
{\sin[(\phi_{j+1}-\phi_{j-1})/2]\sin^2(\phi_j/2)}\right]
^{\Delta p_j^2/4\pi \K}  \nonumber \\ &&~~\times
\exp \left(\frac{1}{4\pi \K}
\sum_{i,j=1 \atop i\neq j}^M\Delta p_i\Delta p_j\ln (1-\cos (\phi_i-\phi_j))\right)\, 
%%\nonumber \\* && ~~~~\times
{\cal K}(\phi_1,\ldots,
\phi_{M-1};\Delta p_1,\ldots,\Delta p_M), %%%\nonumber \\* %%[-6mm] 
\non &&\hspace*{4.5cm}\phi_0=0,
\qquad\phi_{M}=2\pi,
\label{I1}
\end{eqnarray}
where the $\Delta p$'s are particle momenta and
where the ``kernel'' $\cal K$ in general has no momentum dependent 
singularities for $\phi_i\rightarrow
\phi_j$. The first factor in the integrand on the right-hand side is the
well-known Koba--Nielsen integrand, and it produces poles in the
integral for $\phi_i\rightarrow\phi_j$. Physically these poles only
occur when the area law is a good representation of the Wilson loop,
which means that the area should be large and correspondingly the
momenta should also be large. Thus, due to this condition there is no 
tachyon (or other
low lying states) in the spectrum, as one would indeed expect in QCD. We
further show that in the case of large $M$ the kernel $\cal K$ degenerates to a
factor which is essentially independent of  %% the $\phi$'s and 
the $\Delta p$'s.
This implies that when the area behavior of the Wilson loop dominates the
dynamics we
get the interesting result that {\it when many particles are produced
in some collision then the scattering is given by a
Veneziano type of amplitude}. Of course, this is for large $N$ QCD, but
it would be interesting to see to which extent this would be valid
at the LHC collider with $N=3$. 

The condition that the various momentum transfers should be large
in order that the area behavior for the Wilson loop can be inserted
is only a necessary but not sufficient condition for the validity
of our approach. This is important in view of one of the historical reasons
for not using the Veneziano amplitude in strong interaction phenomenology:
at large transverse momenta this amplitude decreases exponentially in 
contrast to the experimental data.
If we consider the 4-point function in terms of the usual
Mandelstam variables $s$ and $t$, our approach is valid when 
$1/\alpha' \la -t\ll s$ in Minkowski space, because it dominates over
other contributions. However, if $-t\sim s$, this is
no longer true, because then the Veneziano amplitude becomes a tiny
exponentially decreasing function and other
contributions lead to a power decrease of the amplitude, which of course
is far more important than  an exponential decrease.  Therefore
the area behaved Wilson loop no longer dominates when $-t\sim s$.

In Sect.~6 we find a relation between the space variables and the external
momenta which is the QCD analogue of the SYM WL/SA duality exhibited in
Eq.~(\ref{du}). If $z(\phi)$ describes the contour of the
Wilson loop as a function of an angular parametrization $\phi$, then
\begin{equation}
z(\phi)=\frac{1}{K}\sum_{i=1}^M\, p_i\,
\Theta (\phi-\phi_i)\,\Theta (\phi-\phi_{i+1}),
\label{analog}
\end{equation}
where $p_i$ are related to the external momenta 
$\Delta p_i$ by $\Delta p_i =p_{i-1}-p_i,$ and $\Theta$ is the usual step
function. If this expression is inserted in the $z$-dependent Wilson loop, it
reproduces the scattering amplitude with a large number $M$ of
external particles when the $\phi_i$'s are integrated over. 
Thus, the relevant
contours are given by the constant momentum vectors, in complete
analogy with Eq.~(\ref{du}). The difference with the supersymmetric case
is that in QCD we have to integrate over the parameters $\phi_i$ which 
represent the points in parameter space where the momenta enter. Otherwise
(\ref{analog}) is like Eq.~(\ref{du}), since $z(\phi)$ equals $p_i/\K$ in the
interval from the point $x_i$ to $x_{i+1}$, so the vector $x_{i+1}-x_i $
equals $\Delta p_i/\K$, as in Eq.~(\ref{du}).  

Some more technical details are discussed in the Appendices.
In Appendix~A we give an example of how to operate with
path integrals related to the ordering of gamma matrices.
In Appendix~B we review the modern approach to the minimal area
as a boundary functional. To illustrate the asymptotic area behavior,
 we evaluate in  Appendix~C the integral over
reparametrizations in the disk amplitude for a large circle.
Appendix~D is devoted to calculations of the path integral over
reparametrizations.
In Appendix~E we consider a scattering amplitude which appears
in QCD, when the Wilson loop is substituted by an exact area law,
i.e.\ \eq{a-l} holds not only asymptotically but for all contours.

\newsection{QCD amplitudes dual to Wilson loops: general results}
%%%{\bf Revised Section 2: QCD amplitudes in terms of the Wilson loops}

In large $N$ QCD, Green's functions of $M$ colorless composite quark
operators (e.g.\ ${\bar q}(x_i) q (x_i)$) are given by the sum
over all Wilson loops passing via the points $x_i$ ($i=1,\ldots,M$), where the
operators are inserted. This approach was first advocated 
on the lattice~\cite{Wil74} and then extended~\cite{MM81} to the continuum.
To obtain a scattering amplitude, one
makes the Fourier transformation with respect to $x_i$ and takes the
corresponding momenta $p_i$ on shell.

The weight for the summation over paths depends on both the quark spin and
the quantum numbers of the operators (see Refs.~\cite{Mig83,Mak02} 
for more detail).
The simplest results are for fermion quarks and the scalar operators
$\bar q (x_i) q (x_i)$ when the connected Green's function  is
\be
G\left(x_1,\ldots,x_M \right) \equiv 
\LA \prod_{i=1}^M \bar q (x_i) q (x_i) \RA_{\rm conn} =
\sum_{C \ni \,x_1,\ldots,x_M} W(C)\,.
\label{ppW}
\ee 
Here $W(C)$ is the Wilson loop in pure Yang--Mills theory at large $N$.
For finite $N$, correlators of several Wilson loops have to be taken into 
account.

In Euclidean space the standard weight for the summation over paths in 
\eq{ppW} reads explicitly
\begin{eqnarray}
&&\hspace*{-5mm}\LA \prod_{i=1}^M \bar q (x_i) q (x_i) \RA_{\rm conn} =
\prod_{i=1}^M  \int\limits_0^\infty \d \tau_i \e^{-m \tau_i}\nonumber \\*
&&\times\int\limits_{z_{i}(0)=x_{i\!-\!1} \atop z_{i}(\tau_i)=x_{i}} \D z_i(t) 
\int\D k(t)~ {\rm sp~P}\exp \left(\i\int\limits_0^{\tau_i} \d t [\dot{z}_i(t)
k(t)-\gamma (t)k(t)]\right)  W(C)\,,
\label{ppWe}
\end{eqnarray}
where the $i$-segment 
of the loop $C$ from $x_{i\!-\!1}$ to $x_i$
is represented by the function $z_i(t)$ ($0<t<\tau_i$)
and $x_M=x_0$ (since the loops are closed). The integration over the
variable gamma matrix $\gamma_\mu (t)$ is explained in Appendix~\ref{appA}. 
Above we have introduced the convention to be used in the rest of the paper
that dot means derivative with respect to whatever argument a function 
has. Thus, for some function $f$ we have $\dot{f}(x)=\d f(x)/\d x$, whereas
$\dot{f}(y)=\d f(y)/\d y$.

Equation~(\ref{ppWe}) is essentially derived in \cite{Mak02}, where further
references can be found. For the readers convenience we repeat the essential 
steps. In QCD the quark fields can be integrated out and we have
the Feynman disentangling\footnote{We use the notation of \cite{Mak02}.
The states $|x\rangle$ etc.\ are eigenstates of $x$ and the
standard Feynman disentangling is used in (\ref{feynman}).}
\begin{eqnarray}
\lefteqn{\LA y|\frac{1}{\gamma_\mu\nabla_\mu +m}|x\RA=\int\limits_0^\infty \d\tau 
\LA y|\e^{-\tau 
(\gamma_\mu \nabla_\mu+m)}|x\RA} \nonumber \\*
&&=\int\limits_0^\infty 
\d\tau \,\e^{-m\tau}\int \D k_\mu~{\rm sp~P}\,\e^{-
\int_0^\tau \d t\, \gamma_\mu (t)(\i k_\mu (t)-\i A_\mu (t))}~\delta(k_\mu (t)+
\i\partial_\mu (t))~\delta(x-y),~~~~ 
\label{feynman}
\end{eqnarray}
where sp and P act on the gammas and color matrices. The functional delta 
function has the representation
\begin{equation}
\delta (k_\mu (t)+\i\partial_\mu (t))~\delta(x-y)=
\int \D v_\mu \,\e^{\i\int_0^\tau 
\d t\,v_\mu (t)(k_\mu(t)+\i\partial_\mu (t))}~\delta (x-y).
\end{equation}
Here $v_\mu(t)$ has no restrictions. Then
\begin{equation}
\delta (k_\mu (t)+\i\partial_\mu (t))~\delta(x-y)=\int \D v_\mu 
\e^{\i\int_0^\tau \d t\, 
v_\mu (t)k_\mu (t)}~\delta (x+\int_0^\tau~\d t\, v(t)-y).
\end{equation}
With $z_\mu (t)=x_\mu +\int_0^t \d t' v_\mu (t')$ we get
\begin{equation}
\delta (k_\mu (t)+\i\partial_\mu (t))~\delta(x-y)=
\int\limits_{z(0)=x \atop z(\tau)=y}\D z_\mu \,
\e^{\i\int_0^\tau \d t \,\dot{z}_\mu (t) k_\mu (t)}.
\label{xxxx}
\end{equation}
Introducing Eq.~(\ref{xxxx}) in Eq.~(\ref{feynman}), we obtain
\begin{eqnarray}
\lefteqn{\int\limits_0^\infty \d\tau \LA y|\e^{-\tau (\gamma_\mu \nabla_\mu+m)}
|x\RA}\nonumber\\*
&&=\int\limits_0^\infty 
\d\tau \,\e^{-m\tau}\int \D k_\mu~\int\limits_{z(0)=x\atop z(\tau)=y}\D z_\mu~
{\rm sp~P}\,\e^{\i\int_0^\tau \d t \,[\dot{z}_\mu k_\mu-\gamma_\mu k_\mu+
\gamma_\mu A_\mu]} \,.
\end{eqnarray}
Shifting the integration by $k_\mu\rightarrow k_\mu+A_\mu$, we get
\begin{eqnarray}
\lefteqn{\int\limits_0^\infty \d\tau \LA y|\e^{-\tau (\gamma_\mu \nabla_\mu+m)}
|x\RA} \non
&& =
\int\limits_0^\infty \d\tau 
\,\e^{-m\tau}\int \D k_\mu\int\limits_{z(0)=x \atop z(\tau)=y}\D z_\mu ~
{\rm sp~P}\,\e^{\i\int_0^\tau \d t\,[\dot{z}_\mu k_\mu -\gamma_\mu k_\mu+ A_\mu 
\dot{z}_\mu]}.
\label{finito}
\end{eqnarray}
This result leads immediately to Eq.~(\ref{ppWe}) since the contour of 
$C$ is composed of pieces each of which can be represented as in
Eq.~(\ref{finito}).

Introducing new proper-time variables
\begin{equation}
\Tau_i=\sum\limits_{j=1} ^i \tau_j \,,\qquad \Tau\equiv \Tau_M
\label{Tau}
\end{equation}
we rewrite \eq{ppWe} in a time-ordered form:
\begin{eqnarray}
\lefteqn{G\left(x_1,\ldots,x_M \right)} \non &&=  \int\limits_0^\infty \d \Tau 
\e^{-m \Tau}
\prod_{i=1}^{M\!-\!1}\int\limits_0^{\Tau_{i\!+\!1}} \d \Tau_i
\int\limits_{z(0)=x_{0} \atop {z(\Tau_i)=x_{i} \atop z(\Tau)=x_M\equiv x_0}} 
\D z(t) \D k(t)
~{\rm sp~P}\,\e^{\i\int_0^{\Tau} \d t\,[\dot{z}(t)k(t)-\gamma (t) k(t)]}\; 
 W[z(t)]\,, \non &&
\label{ppWTau}
\end{eqnarray}
where the loop is represented by a single function 
$z_\mu(t)$ $(0<t<\Tau)$:
\begin{equation}
C \ni \,x_1,\ldots,x_M \;=\,\{z_\mu(t)\,, \quad
z(0)=x_{0} ,\ldots, z(\Tau_i)=x_{i},\ldots, z(\Tau)=x_M\equiv x_0\}\,,
\end{equation}
describing a closed loop passing via the (ordered) set of points $x_i$.
It is essential here that $\Tau_i \leq \Tau_{i\!+\!1}$.

Finally, we introduce the angular-type variables
\begin{eqnarray}
\phi&=&2\pi \frac t\Tau \quad (0<\phi\leq 2\pi) \,, \non
\phi_i&=&2\pi\frac{\Tau_i}\Tau \quad (0<\phi_i\leq 
\phi_{i\!+\!1}\leq 2\pi)\,,
\non
\phi_M&\equiv&2\pi 
\end{eqnarray}
so that
\begin{eqnarray}
\lefteqn{G\left(x_1,\ldots,x_M \right)= \int\limits_0^\infty \d \Tau\, 
\left(\frac\Tau{2\pi}\right)^{M-1}\e^{-m \Tau}} \non
&&
\times\prod_{i=1}^{M\!-\!1}\int\limits_0^{\phi_{i\!+\!1}} \d \phi_i
\int\limits_{z(0)=z(2\pi)=x_{0} \atop z(\phi_i)=x_{i}} \D z(\phi) \int \D
k(\phi)~{\rm sp~P}\,\e^{\i\int_0^{2\pi}\d\phi\, [\dot{z}(\phi)k(\phi)-\Tau
\gamma (\phi)k(\phi)/2\pi]}\; 
 W[z(\phi)]\,.\non
&&
\label{ppWphi}
\end{eqnarray}

The on-shell $M$-particle amplitude can be obtained from the 
Green function~\rf{ppWphi} by applying the standard LSZ reduction
formula. When making the Fourier transformation, it is convenient to
represent $M$ momenta of the (all incoming) particles by the
differences
\be
\Delta p_i= p_{i\!-\!1}-p_i\,.
\ee
Then momentum conservation is automatic while an (infinite) volume 
$V$ is produced, say, by integration over $x_0$. We therefore define
\begin{eqnarray}
G\left(\Delta p_1,\ldots, \Delta p_M \right)= \frac 1V
\prod_{i=1}^M \int {\d^4 x_i} \e^{\i \sum_i \Delta p_i x_i}
G\left(x_1,\ldots,x_M \right)\,.
\label{defmom}
\end{eqnarray}

We can further rewrite this formula, introducing a momentum-space loop
$p_\mu(\phi)$ which is piecewise constant:
\begin{equation}
p(\phi)=p_i \qquad \hbox{for}~ \phi_i<\phi<\phi_{i\!+\!1} \,.
\label{piecewise}
\end{equation}
Noting that 
\be
\dot {\!\!\vec p}(\phi) =- \sum_i \Delta \vec p_i\, \delta(\phi-\phi_i)\,,\qquad
\Delta \vec p_i \equiv \vec p_{i\!-\!1} - \vec p_i\,,
\label{deltas}
\ee
we write
\be
\sum_i \Delta p_i x_i = -\int_0^{2\pi} \d \phi \,\dot p(\phi) \cdot z(\phi)=
 \int_0^{2\pi} \d \phi \,p(\phi) \cdot \dot z(\phi)
\ee
which is manifestly parametric-invariant.

Inserting~\rf{ppWphi} into \eq{defmom} and noting that
\begin{equation}
\prod_{i=1}^M\int\d^4 x_i
\int\limits_{z(0)=x_{0} \atop {z(\phi_i)=x_{i} \atop z(2\pi)=x_0}} \D z(\phi) 
~F(z(\phi ))=
\int\limits_{z(0)= z(2\pi)} \D z(\phi)~ 
F(z(\phi )) \,,
\end{equation}
where $F$ is some functional, we obtain
\begin{eqnarray}
\lefteqn{{G\left(\Delta p_1,\ldots, \Delta p_M \right)=  
\int\limits_0^\infty \d \Tau\, 
\left(\frac\Tau{2\pi}\right)^{M-1}\e^{-m \Tau} }~\prod_{i=1}^{M\!-\!1}
\int\limits_0^{\phi_{i\!+\!1}} \d \phi_i\int \D k(\phi)}\non
&&
\times
\int\limits_{z(0)= z(2\pi)=0} \D z(\phi) ~{\rm sp~P}\,
\e^{\i 
\int_0^{2\pi} \d \phi \,[(k(\phi)+p(\phi)) \cdot \dot z(\phi)-\Tau
\gamma (\phi )k(\phi) /2 \pi]} \,W[z(\phi)]\,,
\label{118}
\end{eqnarray}
where $p(\phi)$ is piecewise constant as is given by \eq{piecewise}. 
We do not integrate over $z(0)=z(2\pi)$ which would produce 
the (infinite) volume factor because of translational invariance. 

In the case of scalar quarks the above procedure can be repeated and
we obtain for the amplitude in position space
\be
G\left(x_1,\ldots,x_M \right) \equiv 
\LA \prod_{i=1}^M \varphi^\dagger (x_i) \varphi (x_i) \RA_{\rm conn} =
\sum_{C \ni \,x_1,\ldots,x_M} W(C)
\label{ppWscalar}
\ee 
or, more explicitly,
\begin{eqnarray}
\lefteqn{
G\left(x_1,\ldots,x_M \right)=  \frac1{2^M} \int\limits_0^\infty \d \Tau\, 
\left(\frac\Tau{2\pi}\right)^{M-1}\e^{-m^2 \Tau/2} }\non
&&~~
\times\prod_{i=1}^{M\!-\!1}\int\limits_0^{\phi_{i\!+\!1}} \d \phi_i
\int\limits_{z(0)=z(2\pi)=x_{0} \atop z(\phi_i)=x_{i}} \D z(\phi) 
\e^{-\frac \mu2 \int_0^{2\pi} \d\phi\,\dot z^2 (\phi)}\; 
 W[z(\phi)]\,,
\label{ppWphiscalar}
\end{eqnarray}
where $\mu=2\pi/\Tau$. In momentum  space this gives
\begin{eqnarray}
\lefteqn{G\left(\Delta p_1,\ldots, \Delta p_M \right)=  
\frac1{2^M} \int\limits_0^\infty \d \Tau\, 
\left(\frac\Tau{2\pi}\right)^{M-1}\e^{-m^2 \Tau/2} }\non
&&~~~
\times\prod_{i=1}^{M\!-\!1}\int\limits_0^{\phi_{i\!+\!1}} \d \phi_i
\int\limits_{z(0)= z(2\pi)=0} \D z(\phi) 
\e^{-\frac \mu2 \int_0^{2\pi}\d\phi\, \dot z^2 (\phi)+ \i
\int_0^{2\pi} \d \phi \,p(\phi) \cdot \dot z(\phi)} \,
W[z(\phi)]
\,.
\label{Gfinscalar}
\end{eqnarray}
In this case there is no $k-$integration and no term involving the gamma 
matrices. The $\dot{z}^2-$term is specific for the scalar case. It is related 
to the occurrence of the second derivative in the Klein-Gordon operator.
Introducing an auxiliary  $k-$integration, we can rewrite \eq{Gfinscalar}
in the form 
\begin{eqnarray}
\lefteqn{{G\left(\Delta p_1,\ldots, \Delta p_M \right)=  
\frac1{2^M} \int\limits_0^\infty \d \Tau\, 
\left(\frac\Tau{2\pi}\right)^{M-1}\e^{-m^2 \Tau/2} }~\prod_{i=1}^{M\!-\!1}
\int\limits_0^{\phi_{i\!+\!1}} \d \phi_i\int \D k(\phi)}\non
&&
~~~\times
\int\limits_{z(0)= z(2\pi)=0} \D z(\phi)\,
\e^{\i 
\int_0^{2\pi} \d \phi \,[(k(\phi)+p(\phi)) \cdot \dot z(\phi)-\Tau
k^2(\phi) /4 \pi]} \,W[z(\phi)]
\label{118s}
\end{eqnarray}
which looks similar to \eq{118}.

\newsection{Wilson loop as string disk amplitude\label{s:3}}

An old idea is to identify the Wilson loop of large $N$ QCD with the
(tree level) disk amplitude in a certain string theory which QCD is 
equivalent to.
As is already pointed out, the simplest Nambu--Goto string in flat space
appears to be surprisingly accurate for large loops and
reproduces the asymptotic area law~\rf{a-l}. 
We shall therefore first review the known
results for the bosonic string.

The calculation of the tree level 
disk amplitude for the Polyakov
string has a subtlety associated with fixing conformal 
gauge~\cite{Alv83,DOP84,CMNP85,Pol87}. The 
decoupling of the Liouville field $\varphi(r,\sigma)$
is possible only in the interior of
the disk, while its boundary value $\varphi(1,\sigma)$ determines the
metric at the one-dimensional boundary:
\be
h(\sigma)=\e^{\varphi(1,\sigma)/2}\,.
\ee
At the classical level this fixes
the parametrization of the boundary contour.
The path integral over the boundary value of the Liouville field then
restores an invariance under reparametrizations of the boundary in quantum
case.

\subsection{Unit circle parametrization\label{ss:uc}}

Let us parametrize the unit disk $\mathbb{D}$ by the variables
$0<r\leq1$ and $\sigma\in [0,2\pi)$. Alternatively, one can map
the unit disk onto the upper half-plane:
\be
z= \i \frac{1+r\e^{\i \sigma}}{1-r\e^{\i \sigma}}\,.
\ee
Then the boundary $\partial\mathbb{D}$,
associated with $r=1$, is mapped onto the real axis parametrized by
$-\infty <s < +\infty$, so that
\be
s(\sigma)= -\cot \frac \sigma2 \,.
\label{themap}
\ee

Integrating over the string fluctuations inside the disk,
i.e.\ over $\vec X(r,\sigma)$ with $r< 1$, we arrive
at the boundary action which is a functional of the field 
\be
\vec X(1,\sigma) \equiv \vec x(\sigma)
\label{b.e.x}
\ee
at the boundary.

Actually the original Fradkin--Tseytlin calculation~\cite{FTs85} used the disk
parametrization.
Their result for the disk amplitude of bosonic string in $d=26$ implies
\be
\Psi[\vec x(\sigma)]= \exp{\Bigg(-\frac \K{2}
\int\limits_{0}^{2\pi} {\d \sigma_1 \d \sigma_2}\,
\vec x(\sigma_1) G^{-1}\left(\sigma_1-\sigma_2\right)\vec x(\sigma_2) 
   \Bigg)}\,,
\label{diskxdisk}
\ee
where 
\be
\K=\frac 1{2\pi\alpha^\prime}
\ee
is the string tension and
\be
G(\sigma)= -\frac{1}{2\pi} \log \left(2-2\cos\sigma \right)=
\frac1\pi \sum_{m=1}^{\infty} \frac{1}{m} \cos m\sigma
\label{Gphi}
\ee
is the corresponding Neumann function for the disk.
It obeys
\be
G^{-1}(\sigma)=-\frac{\d^2}{\d\sigma^2} G(\sigma)=-\frac{1}
{2\pi(1-\cos \sigma)}
=\frac1\pi \sum_{m=1}^{\infty} {m} \cos m\sigma\,.
\label{dotdotG}
\ee

Using \eq{dotdotG} we can rewrite the disk amplitude~\rf{diskxdisk} as
\begin{equation}
\Psi[\vec x(\sigma)]=
 \exp{\Bigg(-\frac \K{2}
\int\limits_{0}^{2\pi} {\d \sigma_1 \d \sigma_2}\,
\dot{\vec x}(\sigma_1) G\left(\sigma_1-\sigma_2\right)\dot{\vec x}(\sigma_2) 
   \Bigg)}\,.
\label{diskxdisk2}
\end{equation}
This formula is pure classical: the exponent is just the classical 
boundary action while determinants coming from the integration over
the fields inside the disk are ignored. For this reason the same result holds
for the bosonic part of the classical boundary action of superstring 
in $d=10$. But more subtle phenomena, such as the L\"uscher term, which
are due to determinants are not captured by \eq{diskxdisk2}.

\subsection{Integration over reparametrizations\label{ss:3.2}}

The exponent in \eq{diskxdisk2} is not invariant under reparametrization
of the contour:
\begin{equation}
\vec x(\sigma)\rightarrow \vec x\left(\theta(\sigma)\right),
\label{reparame}
\end{equation}
where the reparametrizing function $\theta(\sigma)$ obeys
\be
\theta(0)=0\,, \quad \theta(2\pi)=2\pi \,,\quad
\frac{\d\theta(\sigma)}{\d \sigma}\geq 0\,.
\label{tbbc}
\ee

Polyakov~\cite{Pol97} proposed (see also Ref.~\cite{Ryc02}) 
to integrate the disk amplitude~\rf{diskxdisk2} over
the reparametrizations thus providing a reparametrization-invariant
disk amplitude which can be identified with the Wilson loop in large $N$ QCD.
For a unit-circle parametrization this gives
\be
\Psi[\vec x(\cdot)]=
\int\limits_{Diff(S^1)}  \Dp \theta(\sigma) \exp{\Bigg(-\frac \K{2}
\int\limits_{0}^{2\pi} {\d \sigma_1 \d \sigma_2}\,
\vec x\left(\theta(\sigma_1)\right) 
G^{-1}\left(\sigma_1-\sigma_2\right)\vec x\left(\theta(\sigma_2)\right). 
   \Bigg)}
\label{diskxdisktheta} 
\ee
or, equivalently,
\be
\Psi[\vec x(\cdot)]=\int\limits_{Diff(S^1)} 
\Dp \sigma(\theta) \exp{\Bigg(-\frac \K{2}
\int\limits_{0}^{2\pi} {\d \theta_1 \d \theta_2}\,
\dot{\vec x}\left(\theta_1\right)
G\left(\sigma(\theta_1)-\sigma(\theta_2)\right)
\dot{\vec x}\left(\theta_2\right)  \Bigg)}\,.
\label{diskxdisktheta1}
\ee
Here the path integration is over the (infinite) group of diffeomorphisms
of a circle $Diff(S^1)$, i.e.\ over functions with non-negative
derivative $\theta'(\sigma)$, which obey \eq{tbbc}.

An explicit expression for the measure for integrating over 
reparametrizations can be given using an expansion over the complete set
of basis functions $f_j(\sigma)$:
\be
\theta(\sigma)=\sum_{j=1}^L \theta_j f_j(\sigma)\,,
\label{thetaexpand}
\ee
where $f_j=1$ at the $j$-th interval $[\sigma_{j-1},\sigma_j]$ 
($\sigma_0=\sigma_L-2\pi$) with arbitrary $\sigma_j$'s and vanishes otherwise.
A continuous function is approached when the number $L$ of intervals becomes
infinite and $\sigma_{j}-\sigma_{j-1}\to0$. We can then define the measure 
$\Dp\theta(\sigma)$  by 
\be
\int\limits_{Diff(S^1)}  \Dp\theta(\sigma) \cdots
= \lim_{L\to\infty}\int\limits_{0}^{2\pi} 
\d \theta_L \,
\frac{(\sigma_L-\sigma_{L-1})}{|\e^{\i\theta_L}-\e^{\i \theta_{L-1}}|} 
\prod_{j=1}^{L-1}\int\limits_{0}^{\theta_{j+1}} \d \theta_j\,
\frac{(\sigma_j-\sigma_{j-1})}{|\e^{\i\theta_{j}}-\e^{\i \theta_{j-1}}|}  
\cdots \,.
\label{o.i.}
\ee
If $\sigma$ is chosen to be the length of an arc of the unit circle,
then $\sigma_j=2\pi j/L$. Otherwise, we have  
$\sigma_j=\sigma\left(2\pi j/L\right)$.
As distinct from the usual measure $\D \theta(\sigma)$:
%\be
%\int \D\theta(\sigma) \cdots
%= \lim_{L\to\infty}\int\limits_{-\pi}^{+\pi} 
%\prod_{j=1}^{L}{\d \theta_j} \cdots \,,
%\label{o.i.usual}
%\ee
the integrals in \eq{o.i.} are ordered, while the additional factors 
are needed to provide necessary symmetries as is
explained in Sect.~\ref{s:4}.

If the unit circle is mapped onto the real axis by~\eq{themap}, 
the disk amplitude takes the form
\be
\Psi[\vec x(\cdot)]=
\int\limits_{Diff(\mathbb{R})} \Dp t(s) \exp{\Bigg(-\frac \K{4\pi}
\int\limits_{-\infty}^{+\infty} \frac{\d s_1 \d s_2}{(s_1-s_2)^2}
\left[\vec x(t(s_1)) - \vec x(t(s_2)) \right]^2   \Bigg)}\,,
\label{diskx}
\ee
where the path integral over $t(s)$ restores the invariance
under reparametrizations
\be
s\rightarrow t(s)\,,\quad t(-\infty)=-\infty\,,
\quad t(+\infty)=+\infty\,, \quad \frac {\d t}{\d s}\geq 0\,.
\label{reparametrization}
\ee
The measure on $Diff(\mathbb{R})$ can be given by 
\be
\int\limits_{Diff(\mathbb{R})} \Dp t(s) \cdots
= \lim_{L\to\infty}\int\limits_{-\infty}^{+\infty} 
\d t_L \,\frac{(s_L-s_{L-1})}{(t_L-t_{L-1})} 
\prod_{j=1}^{L-1}\int\limits_{-\infty}^{t_{j+1}} \d t_j\, 
\frac{(s_j-s_{j-1})}{(t_{j}-t_{j-1})}  \cdots \,,
\label{o.i.t}
\ee
which is of the same type as in \eq{o.i.}.

The equivalence of \eq{diskx} and \eq{diskxdisktheta} can be shown
using \eq{themap}. For the real-axis parametrization  the Green function 
and  the inverse one read
\be
G(s)= -\frac{1}{\pi} \log |s|
\label{Gt}
\ee
and
\be
G^{-1}(s)=- \frac{1}{\pi s^2}\,.
\label{Gt-1}
\ee
These formulas are the counterparts of Eqs.~\rf{Gphi} and \rf{dotdotG},
while that of \eq{diskxdisktheta1} is
\be
\Psi[\vec x(\cdot)]=
\int\limits_{Diff(\mathbb{R})} \Dp s(t) \exp{\Bigg(\frac \K{2\pi}
\int\limits_{-\infty}^{+\infty} \d t_1 \d t_2\,
\dot{\vec x}(t_1) \log |s(t_1)-s(t_2)|  \dot {\vec x}(t_2)  \Bigg)}\,.
\label{diskx1}
\ee
To show the equivalence of \rf{diskx} and \rf{diskx1} (or \rf{diskxdisktheta} 
and \rf{diskxdisktheta1}), we integrate the exponent by parts 
\bea
\lefteqn{\hspace*{-1cm}
\int\limits_{-\infty}^{+\infty} \frac{\d s_1 \d s_2}{(s_1-s_2)^2}
\left[\vec x(t(s_1)) - \vec x(t(s_2)) \right]^2   =
\int\limits_{-\infty}^{+\infty} 
\frac{\d t_1 \d t_2\, \dot s(t_1) \dot s(t_2)}{[s(t_1)-s(t_2)]^2}
\left[\vec x(t_1) - \vec x(t_2) \right]^2 }\non  
&=&\frac 12 \int\limits_{-\infty}^{+\infty} 
{\d t_1 \d t_2\, }
\left[\vec x(t_1) - \vec x(t_2) \right]^2  
\frac{\partial^2}{\partial t_1 \partial t_2} \log{[s(t_1)-s(t_2)]^2}
%\hspace*{1cm} 
\non
&=& - \int\limits_{-\infty}^{+\infty} 
{\d t_1 \d t_2\, \,\,}
\dot{\!\!\vec x}(t_1)\,\,\dot{\!\!\vec x}(t_2)   
 \log{[s(t_1)-s(t_2)]^2} \,.
\label{rewritex}
\eea

\subsection{Large loops and minimal area}

In spite of the fact that the right-hand side of \eq{diskx} is derivable,
as is already mentioned,
for bosonic string in $d=26$ or superstring in $d=10$, 
where the integration over reparametrizations comes from the integration
over the boundary value of the Liouville field, 
we shall use it only as an ansatz for asymptotically large loops or, 
equivalently, very large $\K$, when the integral over
reparametrizations has a saddle point at $t(s)=t_*(s)$.
This will be crucial for reproducing the area-law behavior
\be
\Psi[\vec x(\cdot)] \stackrel{{\rm large~loops}}\propto 
\e^{-\K S_{\rm min}[\vec x(\cdot)]}
\label{area-law}
\ee
for asymptotically large loops. 

The appearance of the area of the minimal surface, 
spanned by the contour $\vec x(\cdot)$, as a minimum of
the boundary action in \eq{diskx} (or \eq{diskxdisktheta}) is related  
to the fact that it is nothing but the 
functional  known in mathematics as the
Douglas integral~\cite{Dou31}, whose minimum with respect to reparametrizations 
{\em does}\/ give the minimal area: 
\begin{eqnarray}
\lefteqn{{\rm min}_{t(s)} \left\{\frac{1}{4\pi}
\int\limits_ {-\infty}^{+\infty} 
\d s \int\limits_ {-\infty}^{+\infty} \d s^\prime \, 
\frac{\left[x(t(s))-x(t(s^\prime))\right]^2}
{(s-s^\prime)^2} \right\}}\non &&= \frac{1}{4\pi}
\int\limits_ {-\infty}^{+\infty} 
\d s \int\limits_ {-\infty}^{+\infty}  \d s^\prime \, 
\frac{\left[x(t_*(s))-x(t_*(s^\prime))\right]^2}
{(s-s^\prime)^2}= S_{\rm min}\left[x(\cdot)\right].
\label{Ds}
\end{eqnarray}
This issue is clarified in Appendix~B. 
%%% as well as in the Appendices to Refs.~\cite{Mig93,Ryc02}.

The necessity of the function  $t_*(s)$ (or $\theta_*(\sigma)$),
reparametrizing the boundary, is due to the fact that coordinates 
describing the minimal surface have to obey the condition of the 
conformal gauge for the quadratic action, used in the Polyakov
string formulation, to coincide with the area. 
In particular, \eq{b.e.x} has to be replaced  for this reason by 
\be
\vec X(1,\sigma) \equiv \vec x(\theta_*(\sigma)) \,.
\label{b.e.x1}
\ee 

For large loops $\vec x(\cdot)$
the path integral over reparametrizations in Eqs.~\rf{diskx} or \rf{diskx1}
has a saddle point,
which is denoted above by $\theta_*(\sigma)$. The saddle-point value of the
boundary action recovers the minimal area $S_{\rm min}[\vec x(\cdot)]$,
reproducing the exponential in \eq{area-law}. To say in other words,
minimizing the exponential of the Douglas functional over reparametrizations
yields the precise area-law
\be
{\rm min}_{t(s)} \left\{\e^{- \frac{\K}{4\pi}\int\limits_ {-\infty}^{+\infty} 
\d s \int\limits_ {-\infty}^{+\infty} \d s^\prime \, 
\frac{\left[x(t(s))-x(t(s^\prime))\right]^2}
{(s-s^\prime)^2}} \right\} = \e^{-\K S_{\rm min}\left[x(\cdot)\right]}.
\ee 
We shall use this equation in Sect.~\ref{s:5} when calculating
scattering amplitudes in QCD.

The Gaussian fluctuations around the saddle-point $\theta_*(\sigma)$
result in an appearance of a pre-exponential factor, so
\eq{area-law} is modified as 
\be
\Psi\left[x(\cdot)\right] \stackrel{{\rm large~loops}}=
F\left[\sqrt{\K}x(\cdot)\right]
\e^{-\K S_{\rm min}\left[x(\cdot)\right]}
\left[1+ {\cal O}\left((K S_{\rm min})^{-1}\right)\right]\,,
\label{circle1}
\ee
%\be
%\Psi[{\rm circle}] \stackrel{{\rm large}\;R}\propto 
%({\K} R^2)^{1/2}
%\e^{-\pi \K R^2}\left[1+ {\cal O}\left((KR^2)^{-1/2}\right)\right]\,,
%\label{circle1}
%\ee
%For the contour in the form of
%a circle of the radius $R$, when $\theta_*(\sigma)=\sigma$ as is 
%shown in Appendix~B, the amplitude  $\Psi[x(\cdot)]$ depends only on
%$K R^2$ and for large $K R^2$ has the structure
%where the constant $C$ is finite  %%% and contour-dependent 
%as is shown in Appendix~C, where it is also discussed the case
%of an arbitrary contour, when the area-law~\rf{area-law} is reproduced
%for large contours with additional finite but contour-dependent 
%pre-exponential. 
where the pre-exponential factor 
$F\left[\sqrt{\K}x(\cdot)\right]$ is contour dependent.
Its calculation for a circle  is performed in Appendix~C.

Therefore, the asymptotic area law~\rf{area-law} is recovered
by Eqs.~\rf{diskxdisktheta} [\rf{diskxdisktheta1}] or \rf{diskx}
[\rf{diskx1}] modulo the pre-exponential which is not essential for large 
loops.

\subsection{Remark on the area law for large loops in QCD}

In general, the Wilson loops are not observable in QCD.
Observable quantities are scattering amplitudes which are given
by the sum over paths of the Wilson loops (see \eq{ppWphi}).
Nevertheless, some observables can be directly expressed through
the Wilson loop of a certain shape. An example is the interaction potential
between static quarks which is determined by a $T\times R$ rectangular
loop for $T\gg R$. 

It is well-known~\cite{Alv81,Arv83,Orl01}
how this potential is calculable for the Nambu--Goto string, including
the L\"uscher term. 
Then the most convenient choice is to parametrize the string world-sheet
by the coordinates along the $T$- and $R$-axes, which are
conformal. The results described in the previous subsection allow us to
answer a natural question as to how the linear potential 
can be reproduced for the unit-circle 
parametrization which does {\em not}\/ obey in general the conformal gauge.
The reparametrizing function $\theta_*(\sigma)$
is determined for the rectangle loop by the Schwarz--Christoffel mapping.
Once again, the path integral over reparametrizations in \eq{diskx} is 
{\em crucial}\/
to identify the asymptote of the string disk amplitude with
the asymptotic area-law behavior of the Wilson loop in QCD. 

\newsection{Derivation of the Koba--Nielsen amplitudes\label{s:4}}

Scattering amplitudes in open string theory are conventionally obtained
by inserting vertex operators in the path integral over string fluctuations.
Having the boundary action, they are represented as the path integral
over $\vec x(\cdot)$ with the vertex operators inserted
at the boundary. The disk
amplitude plays, therefore, the role of a generating functional for
scattering amplitudes. 

\subsection{The Fourier transformation\label{ss:ft}}

%%For the reasons to become immediately clear, 
It is convenient to perform a (functional) Fourier transformation%
\footnote{Such a transformation of the Wilson loops  
was first advocated in Ref.~\cite{Mig86}.}
\be
\Psi[\vec p(\cdot)]= \int \D \vec x \e^{\i \int \vec p \,\d\vec x }
\Psi[\vec x(\cdot)]\,,
\label{Fourier}
\ee
where the exponent
\be
\int  \vec p \,\d \vec x = \int \d t \,\vec p \,\dot {\vec x}  
\ee
is invariant under reparametrizations. This functional Fourier transformation 
is of the type as in \eq{118} and transforms the position-space disk 
amplitude to a momentum-space one.

Substituting~\rf{diskx} into \eq{Fourier} and performing the Gaussian
integration, we arrive at the following momentum-space disk amplitude
\be
\Psi[\vec p(\cdot)]=
\int \Dp t(s) \exp{\Bigg(-\frac {\alpha^\prime}2
\int\limits_{-\infty}^{+\infty} \frac{\d s_1 \d s_2}{(s_1-s_2)^2}
\left[\vec p(t(s_1)) - \vec p(t(s_2)) \right]^2   \Bigg)}\,.
\label{diskp}
\ee
Note that \rf{diskp} has the same form as \rf{diskx} only with 
$\K$ replaced by $1/\K=2\pi\alpha^\prime$.

In front of the exponential in \eq{diskp} there is in fact the
determinant (in $d=4$ dimensions) 
\bea
\left[\det_{t_1 t_2} G\left(s(t_1)-s(t_2)\right) \right]^{-d/2}&=&
\exp\left\{-\frac d2\int \d t_1 \log [ G\left(s(t_1)-s(t_2=t_1)\right)] \right\}
\non &=&
\exp\left\{-\frac d2\int \d t_1 \log [ G(0)] \right\}\,,
\eea
which is an infinite constant. It can be regularized for the 
unit-circle parametrization by modifying
the Green function~\rf{Gphi}:
\be
G_b(\sigma)= -\frac{1}{2\pi} \log \left(1+b^2-2b\cos\sigma \right)=
\frac1\pi \sum_{m=1}^{\infty} \frac{b^m}{m} \cos m\sigma \,,
\label{Gphib}
\ee
which yields 
\be
\left[\det_{\theta_1 \theta_2} 
G\left(\sigma(\theta_1)-\sigma(\theta_2)\right) \right]^{-d/2}=
\exp\left\{-d \pi \log [ G_b\left(0 \right)] \right\}\,.
\ee
What is important is that this regularized determinant does not
depend on the reparametrizing function $\sigma(\theta)$ and 
therefore is an overall constant.

The main advantage of the momentum-dependent amplitude~\rf{diskp} is that
the momentum variable can be chosen to be a step function of $t$:
\be
\vec p (t)= \sum _{j=1}^M \vec p_j f_j(t)\,,
\label{genexpansion}
\ee
where $f_j=1$ at the $j$-th interval $[t_{j-1},t_j]$ and vanishes otherwise
as in \eq{thetaexpand}.
Note that the stepwise discretization of $\vec x(t)$ itself is not possible
since it would violate the continuity of the world-line of the string end.

Since $\vec p(t)=\vec p_j$ at the $j$-th interval, the exponential in
\eq{diskp} is in fact a function of $M$ variables $s_j$: $s_j<s_{j+1}$.
The only effect of the reparametrization~\rf{reparametrization} is 
then to change the values of $s_j$'s 
keeping their cyclic order:
\be
\{s_1 < \ldots< s_{j-1}< s_j<\ldots< s_M \} \to
\{t_1< \ldots< t_{j-1}< t_j<\ldots< t_M \} \,.
\label{discretereparametrization}
\ee
This is a discrete version of the transformation~\rf{reparametrization}.

%%%In \eq{diskp} we can 

\subsection{From disk amplitude to Koba--Nielsen amplitude\label{ss:4.2}}

The stepwise discretization of the momentum-space loop from the previous 
subsection naturally results in the Koba--Nielsen amplitudes.
We present here a simple on-shell derivation which is pretty close
to the standard derivations in string theory.

Integrating by parts as in \eq{rewritex}, we rewrite the exponent in 
\rf{diskp} as
\be
%%\bea
%%\lefteqn{\hspace*{-1cm}
\int\limits_{-\infty}^{+\infty} \frac{\d s_1 \d s_2}{(s_1-s_2)^2}
\left[\vec p(t(s_1)) - \vec p(t(s_2)) \right]^2   
%%=\int\limits_{-\infty}^{+\infty} 
%%\frac{\d t_1 \d t_2\, \dot s(t_1) \dot s(t_2)}{[s(t_1)-s(t_2)]^2}
%%\left[\vec p(t_1) - \vec p(t_2) \right]^2 }\non  
%%&=&\frac 12 \int\limits_{-\infty}^{+\infty} {\d t_1 \d t_2\, }
%%\left[\vec p(t_1) - \vec p(t_2) \right]^2  
%%\frac{\partial^2}{\partial t_1 \partial t_2} \log{[s(t_1)-s(t_2)]^2}
%%%\hspace*{1cm} 
%%\non&=& 
=- \int\limits_{-\infty}^{+\infty} 
{\d t_1 \d t_2\, \,\,}
\dot{\!\!\vec p}(t_1)\,\,\dot{\!\!\vec p}(t_2)   
 \log{[s(t_1)-s(t_2)]^2} \,.
\label{rewrite}
%%\eea
\ee
For the step functions~\rf{genexpansion}, when
\be
\dot {\!\!\vec p}(t) = -\sum_j \Delta \vec p_j \,\delta(t-t_j)\,, \qquad
\Delta \vec p_j \equiv \vec p_{j-1}- \vec p_j \,,
\label{deltas1}
\ee
we find 
\be
\Psi(\Delta p_1,\ldots,\Delta p_M)= 
\prod_i (1-b)^{\alpha^\prime \Delta \vec {p}_i^2}
\int\limits_{\t_1<\ldots<\t_{j-1}<\t_j<\ldots<\t_M} \prod_j \d \t_j
\prod_{k\neq l} |\t_k-\t_l|^{\alpha^\prime\Delta \vec p_k\Delta \vec p_l }\,.
\label{onshell}
\ee

A few comments concerning \eq{onshell} are in order.
The factor in front of the integral comes from the multipliers with $k=l$.
They are made finite by a regularization of the type in \eq{Gphib}.
The integration over $\t_j$'s results from the integration over
$\t(\s_j)$ in the path integral over reparametrizations in \eq{diskp},
while the integration over $\t(\s)$ at the intermediate points is naively
ignored. This integration is in the spirit of the integration over
the positions of vertex operators in string theory.
Setting $\alpha^\prime \Delta \vec p_k^{\,2}=1$ (i.e.\ imposing the 
tachyonic on-shell condition) and fixing
the $PSL(2;\mathbb{R})$ invariance, appearing after this setting, 
in the standard way,
we obtain the on-shell tachyonic amplitude in the Koba--Nielsen variables.
%%which gives linear Regge trajectories. %%% $\alpha(t)=1+\alpha' t$.

\subsection{Projective-invariant off-shell amplitudes\label{ss:4.3}}

As is well-known, the amplitude~\rf{onshell} is invariant under 
a transformation from the
$PSL(2;\mathbb{R})$ projective linear group
\begin{equation} 
\t\rightarrow \t^\prime=\frac {a \t+b}{c \t+d}\,, \qquad a d-b c=1
\label{projective}
\end{equation}
only when  $\alpha^\prime\Delta \vec p_k^{\,2} =1$, \ie only for tachyonic 
amplitudes. On the contrary we might expect that an $M$-particle amplitude,
generated by the Fourier transformation of the reparametrization-invariant
momentum-space disk amplitude~\rf{diskp}, should be projective-invariant
because the projective group is a subgroup of reparametrization
transformations.

There are two reasons why~\rf{diskp} resulted in \eq{onshell}.
The first reason is the divergence of the double integral 
over $s_1$ and $s_2$ in the exponent in \eq{diskp} for $s_1=s_2$. 
This integral is of course
convergent for smooth $p(t(s))$, but it is divergent for the stepwise   
$p(t(s))$ when $s_1$ and $s_2$  lie on adjacent sides $k=l\pm 1$. 
The correct procedure is to understand this
integral according to the principal-value prescription which lead us
to the prescription to omit the adjacent sides
with $k=l\pm1$, as is shown in Appendix~D.

If we repeat the calculation omitting the sides
with $k=l\pm1$, then  the integrations over $s_1$ and $s_2$ are
perfectly finite resulting in
\bea
\lefteqn{\sum_{k\neq l\pm1}\; \int\limits_{\t_{k-1}}^{\t_k} \d \t_1
\int\limits_{\t_{l-1}}^{\t_l} \d \t_2\, \frac{(p_k-p_l)^2}{(\t_1-\t_2)^2} } \non
&=& -2\sum_{k\neq l}\; \Delta p_k \cdot\Delta p_l \log |\t_k-\t_l|
-2 \sum_j \Delta p_j^2 \log \frac{(\t_j-\t_{j-1})(\t_{j+1}-\t_j)}
{(\t_{j+1}-\t_{j-1})}
\label{piip}
\eea
which is projective invariant.

The second reason for the loss of the projective invariance in \eq{onshell}
is that the integration over reparametrizations at intermediate points of
a side was not taken into account. A rather
subtle issue how to accurately integrate over $\t(\s)$ at the intermediate 
points is explained in Appendix~D, 
where it is shown that this integration 
%%The main integral is
%%\be
%%\int \Dp t(s) \, \prod_{i=1}^M \delta\left( t(s_j)-t_j   \right)=
%%\prod_{i=1}^M \frac{1}{|t_j-t_{j-1}|}\,\Theta(t_j-t_{j-1}) \,,
%%\label{master}
%%\ee
over reparametrizations at the intermediate points 
results in the following measure 
\be
D^{(M)} \t = \prod_{j=1}^M \frac{\d \t_j}{|\t_j-\t_{j-1}|} 
%%%\Theta(\t_j-\t_{j-1}) 
\label{measure11}
\ee  
for the integration over $\t_j$'s. 
It is invariant under the projective 
transformation~\rf{projective} and gives
\begin{eqnarray}
\lefteqn{\Psi(\Delta p_1,\ldots,\Delta p_M)}\non &=& \!\!\!\!\!\!
\int\limits_{\t_1<\ldots<\t_{i-1}<\t_i<\ldots<\t_M} \!\!\prod_i\frac{\d \t_i}
{|\t_i-\t_{i-1}|} 
\prod_{k\neq l} |\t_k-\t_l|^{\alpha^\prime\Delta \vec p_k\Delta \vec p_l }
\prod_j\left(\frac{(\t_j-\t_{j-1})(\t_{j+1}-\t_j)}{(\t_{j+1}-\t_{j-1})}\right)
^{\alpha^\prime \Delta p_j^2} \!.\non &&
\label{simple}
\end{eqnarray}

For the  case of 4 scalars, \eq{simple} reproduces the Veneziano
amplitude
\be
A(\Delta p_1,\Delta p_2,\Delta p_3,\Delta p_4)
=\int\limits_0^1 \d x\, x^{-\alpha(s)-1}
(1-x)^{-\alpha( t)-1} \,,
\label{veneziano}
\ee
where $\alpha(t)=\alpha^\prime t$ and
\bea
s&=&-(\Delta p_1+\Delta p_2)^2=-(\Delta p_3+\Delta p_4)^2 \,,\non
t&=&-(\Delta p_2+\Delta p_3)^2=-(\Delta p_1+\Delta p_4)^2
\eea
are usual Mandelstam's variables (for Euclidean metric).
Here the tachyonic condition $\alpha^\prime \Delta p_j^2=1$
has not  to be imposed.

\subsection{Ambiguities of the measure}

In Ref.~\cite{MO08} we used, instead of \rf{measure11}, 
another choice  of the measure 
\be
D ^{(M)} \t= 
\prod_{i=1}^{M} \d \t_i 
\frac{(\t_{i+1}-\t_{i-1})}{(\t_i-\t_{i-1})(\t_{i+1}-\t_i)}
\label{measure}
\ee
which is also invariant under the projective transformation~\rf{projective}.
This results in the amplitude
\begin{eqnarray}
\lefteqn{\Psi(\Delta p_1,\ldots,\Delta p_M)}\non &=& 
\int\limits_{\t_1<\ldots<\t_{i-1}<\t_i<\ldots<\t_M} \prod_i \d \t_i
\prod_{k\neq l} |\t_k-\t_l|^{\alpha^\prime\Delta \vec p_k\Delta \vec p_l }
\prod_j\left(\frac{(\t_j-\t_{j-1})(\t_{j+1}-\t_j)}{(\t_{j+1}-\t_{j-1})}\right)
^{\alpha^\prime \Delta p_j^2-1}
\,. \non &&
\label{Lovelace}
\end{eqnarray}
These formulas are known as the Lovelace choice~\cite{Lov70}, that
reproduces some off-shell dual amplitudes known since late 1960's
(for a review see Ref.~\cite{DiV92}, references therein and 
subsequent papers~\cite{CLP98,LPM99}).
For the case of 4 scalars, \eq{Lovelace} reproduces the Veneziano
amplitude~\rf{veneziano}
with $\alpha(t)=1+\alpha^\prime t$. It has the same intercept of the 
Regge trajectory as the on-shell amplitude~\rf{onshell}
but now the tachyonic condition $\alpha^\prime \Delta p_j^2=1$
has not to be imposed. 

The measures~\rf{measure11} and \rf{measure} can be generalized as
\be
D^{(M)} \t= \prod_i \frac{\d \t_i }{(\t_i-\t_{i-1})}\,
\left(\frac{\t_{i+1}-\t_{i-1}}{\t_{i+1}-\t_i}\right)^{\alpha_0}\,,
\label{measurea}
\ee
where $\alpha_0$ is an arbitrary constant. The measures~\rf{measure11} and
\rf{measure} are reproduced for $\alpha_0=0$ and $\alpha_0=1$, respectively. 
The measure~\rf{measurea} is in the spirit of the
Koba--Nielsen amplitudes and is projective invariant for an arbitrary 
$\alpha_0$.   It is the measure~\rf{measure11} which we have obtained 
in the previous subsection by integrating over reparametrizations, but
we cannot exclude that a similar procedure may also exist for other measures.

The amplitude associated with the measure~\rf{measurea} is again the
 Veneziano amplitude~\rf{veneziano} with $\alpha(t)=\alpha_0+\alpha^\prime t$.
Therefore, only an intercept of the Regge trajectory is sensitive to the choice 
of the measure provided it is projective invariant.
But the change of the measure~\rf{measurea} becomes not essential for 
$1/\alpha^\prime \ll -t \la s$,
when the integral over $x$ is dominated by a saddle point.

The Regge trajectory with $\alpha(0)\approx1$ is usually associated
with the vacuum trajectory = Pomeron which 
appears in QCD from cylinder diagrams $\sim 1/N^2$ (with two quark
loops). We work with planar diagrams, single quark loop and, correspondingly, 
with a quark-antiquark Regge trajectory which has 
$\alpha(0)\approx 1/2$ from experiment. 
%%Is it possible to have   $\alpha(0)=1/2$  for the
%%Neveu-Schwartz dual model which may come from
%%the superstring boundary action?

\newsection{QCD amplitudes and the area behavior of W\label{s:5}}

In the following we ask what is the contribution from the area behavior of the 
Wilson loop to the corresponding QCD amplitudes. It is of course clear that
depending on the kinematical situation there will be other important 
contributions, e.g.\ at large transverse momenta where a perturbative
behavior of $W$ is relevant.

As discussed in the introduction it is well-known from lattice calculations 
that the area behavior, and
more generally large distance results, have been found to follow the 
predictions from the Nambu--Goto action rather precisely even at 
surprisingly low distances. It therefore makes sense to ask what
happens to the QCD amplitudes if we approximate the Wilson loop by the area 
behavior. This question will be studied in the present section.

We now want to insert the Wilson loop given by the area behavior,
taking the latter from the celebrated Douglas construction \cite{Dou31}
discussed in details in Sect.~\ref{s:3} and Appendix~B. We write 
\begin{equation}
W={\cal SP}_{\sigma}~\int \D\sigma (\tau) \exp \left(\frac{K}{4\pi}
\int_0^{2\pi}
\d\theta\int_0^{2\pi}\d\theta' \dot{z}(\theta)\dot{z}(\theta')\ln (1-
\cos [\sigma (\theta)-\sigma (\theta')])\right).
\label{D7}
\end{equation}
Here $\cal SP_\sigma$ means that the full integral should be 
replaced by the {\it Saddle Point}\/ with respect to
the reparametrizations $\sigma (\phi)$, where $\sigma$ satisfies 
$\dot{\sigma}>0$ and $\sigma (0)=0$ and $\sigma (2\pi)=2\pi$. The integral
over $\sigma (\tau )$ would then  produce the exponential of the minimal 
area times a prefactor as displayed in \eq{circle1}. 
However, since we want to consider QCD, then only the
minimal area factor is pertinent, as we know from lattice gauge calculations.
In QCD there are subleading factors (perimeter term, L\"uscher term, ...), but
they are probably not given by the prefactor coming from the saddle
point in (\ref{D7}). Therefore the $\cal SP_\sigma$ is restricted to include
only the leading saddle-point contribution with no multiplicative
prefactors.

In the following we want to insert Eq.~(\ref{D7}) in the basic formula
(\ref{118}) with the goal of doing the $z-$integral. If we perform the
Douglas saddle point $\cal SP_\sigma$ first, then the resulting minimizing 
functions 
$\sigma$ will depend on the curves $C$, and hence also on parameters entering
in $z$. As an example, in Appendix~B 
we have discussed the elliptic case, where
the minimizing $\sigma$ depends on the ratio of the lengths of the two axes,
and these lengths also enter in the expression for $z$.
Hence, following this procedure the $z-$integral cannot be performed as
a simple Gaussian integral. Therefore, in applying Eq.~(\ref{118}) we
shall assume that the $\cal SP_\sigma$ operation commutes with the $z-$integral
entering in (\ref{118}), 
\begin{equation} 
\int\D k(\tau) \int \D z(\tau)~{\cal SP}_\sigma\int \D \sigma (\tau)=
\int\D k(\tau)~ {\cal SP}_\sigma\int \D \sigma (\tau)\int \D z(\tau)\,,
\label{assumption}
\end{equation}
so that finding the saddle point  can wait until the $k-$integral is to be 
performed.%
\footnote{It worth noticing that the $\cal SP_\sigma$
operation is in fact nothing but taking the classical limit $\hbar\to0$. 
If the dependence on
Planck's  constant  $\hbar$ is restored, the exponents in  both the
coordinate-space disk amplitude~\rf{diskx} and the momentum-space disk
amplitude~\rf{diskp} are divided by  $\hbar$ because it enters the exponent
of the Fourier transformation.} 
The validity of Eq.~(\ref{assumption}) is plausible {\it a posteriori},
since the $\sigma-$dependence in all situations encountered turns out to 
occur entirely in the logarithm
$ %%\begin{equation}
\ln (1-\cos (\sigma (\tau) -\sigma (\tau'))),
$ %%\end{equation}
as we shall see in the following.  

With this assumption we can now
perform the Gaussian $z-$integration in the basic formula (\ref{118})
to obtain 
\begin{eqnarray}
&&G\left(\Delta p_1,\ldots, \Delta p_M \right)= 
\int\nolimits_0^\infty \d \Tau\,\e^{-m \Tau} %%\non &&\times
\int\nolimits_0^{\Tau} \d \tau_{M-1}
\prod_{i=1}^{M\!-\!2}\int\nolimits_0^{\tau_{i\!+\!1}} \d \tau_i\,
\int\D k(\tau)~ {\cal SP}_\sigma\int \D \sigma (\tau)\nonumber \\ 
&& ~\times {\rm sp~P}\,
\e^{\frac {\alpha^\prime}{2}\int_0^{\Tau} \d\tau \int_0^{\Tau} \d\tau' 
(\dot{k}(\tau)+\dot{p}(\tau))\cdot  (\dot{k}(\tau')+\dot {p}(\tau')) 
\ln\left(1-\cos (\sigma (\tau)-\sigma (\tau'))\right)
-\i\int_0^{\Tau}\d\tau\, \gamma (\tau)\cdot k(\tau)}.~~~~
\label{sves1}
\end{eqnarray}
Here it should be emphasized that the dots in Eq.~(\ref{D7}) can be moved
to the logarithm by partial integrations, and hence the integration over
$z$ can be formulated such that it involves only $z$ itself and not $\dot{z}$.
We mention that a similar simplification does not occur 
in Eq.~(\ref{Gfinscalar}) for scalar quarks, 
due to the occurrence of $\dot{z}^2$.

Inserting the stepwise ${p}(\tau)$, regularizing the integral in the exponent
by the principal-value prescription and using the formula of the type of
\eq{piip}, this becomes\begin{eqnarray}
\lefteqn{G(\Delta p_1,...,\Delta p_M)\propto 
\prod_1^{M-1}\int\limits_0^{\phi_{i+1}}\d\phi_i\, 
\prod_{j=1}^M
\left[\frac{\sin[(\phi_{j+1}-\phi_j)/2]\sin[(\phi_{j}-\phi_{j-1})/2]}
{\sin[(\phi_{j+1}-\phi_{j-1})/2]\sin^2(\phi_j/2)}\right]
^{\Delta p_j^2/4\pi \K} } \non &&\times
\exp \left(\frac{1}{4\pi \K}
\sum_{i,j=1 \atop i\neq j}^M\Delta p_i\Delta p_j\ln (1-\cos (\phi_i-\phi_j))\right)\, 
%%\nonumber \\* && ~~~~\times
{\cal K}(\phi_1,\ldots,
\phi_{M-1};\Delta p_1,\ldots,\Delta p_M), 
\non &&\hspace*{4.5cm} \phi_0=0,
\qquad\phi_{M}=2\pi,
\label{L1}
\end{eqnarray}where the ``kernel'' $\cal K$ is given by
\begin{eqnarray}
{\cal K}&=&\int \D k(\theta)\,{\cal SP}_\sigma \int \D\sigma 
(\phi)  \non &&\times\exp\left(\frac{1}
{4\pi K}\int_0^{2\pi} \d\theta\int_0^{2\pi} \d\theta'\,
\dot{k}(\theta)\dot{k}(\theta')\ln (1-\cos (\sigma (\theta)-\sigma (\theta')))
\right)\nonumber \\
&&\times\exp\left(\frac{1}{2\pi K}\sum_i\Delta p_i\int_0^{2\pi} \d\theta\,
\dot{k}(\theta)\ln (1-\cos (\phi_i-\sigma(\theta))\right)\nonumber \\
&&\times \int\limits_0^\infty \d\tau~\tau^{M-1}\,\e^{-m\tau}~ {\rm sp~P}\,\exp 
\left(-\frac{\i\tau}{2\pi}
\int_0^{2\pi}\d\phi\, \gamma (\phi)k(\phi)\right).
\label{L2}
\end{eqnarray}
Minimizing the kernel $\cal K$ with respect to $\sigma$ gives the requirement 
\begin{equation}
\int\limits_0^{2\pi} \d\theta' \,\dot{k}(\theta)\dot{k}(\theta')
\cot{\frac{\sigma_*(\theta)-\sigma_* (\theta')}{2}}
+2\sum_i\Delta p_i~\dot{k}(\theta)\cot
\frac{ \phi_i-\sigma_* (\theta)}{2}=0.
\label{L3}
\end{equation}
For a given $k$ this determines $\sigma_* (\theta)$ or, alternatively, the 
minimizing reparametrization $\theta_* (\sigma)$. Using $\sigma_*$
it follows by our assumption (\ref{assumption}) that only the minimal areas  
are included as contributions to the amplitude. 

The main result (\ref{L1}) reveals the interesting appearance of
poles of the Veneziano type. These occur when two $\phi$'s coincide, 
$\phi_i\rightarrow \phi_j$. The kernel $\cal K$ does not in general have any 
pole singularities dependent on the momenta when $\phi_i\rightarrow \phi_j$, 
so the momentum dependent Veneziano-type poles 
cannot be canceled by any contribution from $K$. 

The factor
\begin{equation}
\exp \left(\frac{1}{4\pi K}
\sum_{i\neq j}\Delta p_i\Delta p_j\ln (1-\cos (\phi_i-\phi_j))\right)
\label{L4}
\end{equation}
is independent of the reparametrizations involved in the Douglas construction.
This is due to the fact that $\dot{p}$ is a simple sum of delta functions.
The above factor is therefore {\it universal}. 

After  a  transformation of the variables $\phi_i$, Eq.~(\ref{L4}) becomes
similar to the Koba--Nielsen representation of the $M$-point function. 
To see this use%
\footnote{Here and below the value of $\phi_M$ can be chosen arbitrary, 
respecting the cyclic symmetry.}
\begin{equation}
\ln \left[2(1-\cos (\phi_i-\phi_j))\right]
=2 \ln \left|2\sin((\phi_i-\phi_j)/2)\right|=
2 \ln |\t_i-\t_j|+\ldots,\quad \t_i=-\cot (\phi_i/2),
\end{equation}
where we left out terms that vanish in the sums occurring  in
Eq.~(\ref{L4}) due to energy-momentum conservation. The $\t_i$'s occur like 
the variables in the Koba--Nielsen formula.\footnote{The differentials 
transform as $\d\phi_i=2\d \t_i/(1+\t_i^2)$.} 
Thus, in QCD the angular variables, related to 
the proper-time variables by $\phi_i=2\pi \tau_i/\tau$, play the role
of the Koba--Nielsen variables.

We thus see that the usual dual model poles are present in the QCD
amplitude (\ref{L1}). This is indeed to be expected on intuitive grounds,
because the area behavior of the Wilson loop can be interpreted in a string
framework as arising from the rotating stick with a well-known Regge
type spectrum.%
\footnote{Another derivation of the Regge spectrum from the are law was
given in Ref.~\cite{DKS94}.} It is therefore quite satisfactory that this 
result also occurs in our general QCD formula when the area law is imposed.

It must be emphasized that in QCD the area law is valid only for large areas, 
i.e.\ for large distances. Translating this to momentum space, we need large 
momenta. Therefore, the low lying dual model spectrum is not relevant
in QCD. Thus, the tachyon is also of no relevance, as should be the case.

Going back to Eq.~(\ref{L1}), it should be noted that this formula does not
correspond to a simple dual amplitude, because of the additional 
factor $\cal K$.
Thus, although the spectrum is quite stringy, the amplitude is more
complicated than in the standard dual models. Presumably, this is not
too surprising.

As is mentioned above, the tachyon does not occur due to the stringy 
correspondence 
large distances=large momenta (in contrast to perturbation theory).
Also, when the area behavior leads to extremely small contributions, as is the 
case for large transverse momenta in string amplitudes (for the
4-point function $-t \sim s$), these are not 
important relative to perturbative contributions%
\footnote{We would like to emphasize once again that we are dealing in
the large $N$ limit with the quark-antiquark Regge trajectory,
whose perturbative QCD calculation was pioneered in Ref.~\cite{KL83}.}, 
which would then dominate the Wilson loop.
This is how the exponential falloff of the 
4-particle amplitude with large
$-t\sim s$, which is unavoidable in string theory~\cite{GM87}, does
not happen in our consideration. 

Now let us ask if it is possible (at large momenta) to obtain something
like a standard dual model. This depends on the extra factor $\cal K$ in 
Eq.~(\ref{L1}). The last factor in the integral over $k$ in (\ref{L2})
suggests that we substitute
\begin{equation}
k\rightarrow  k /\tau.
\end{equation}
Then we see that the first two factors in the definition of $\cal K$ become
close to one if $\tau$ is large. For the number of external mesons
$M$ large, the $\tau-$integral in Eq.~(\ref{L2}) is dominated by large
\begin{equation}
\tau =(M-1)/m \,.
\label{domi}
\end{equation}
It is seen from this formula that
the values of $\tau$, dominating the $\tau-$integral in Eq.~(\ref{L2}),
are also large for small values of $m$, but we consider such 
a limit to be rather
formal because $m$ has the meaning of a constituent quark mass in QCD,
which is about hundred MeV from experiment even for very small bare masses 
of up and down quarks because of the spontaneous breaking of chiral symmetry.

When the large values of $\tau$ \rf{domi} dominate,
the integrand of the integral over $\sigma(\phi)$ in \eq{L2}
does not depend on $\sigma$ and ${\cal K}$ degenerates into 
\be
{\cal K}\propto \int\limits_{\sigma(\phi_i)=\phi_i} \D\sigma(\phi) =
\prod_{i=1}^M
\frac{\sin(\phi_{i+1}/2)\sin(\phi_i/2)}{\sin[(\phi_{i+1}-\phi_i)/2]} 
=\prod_{i=1}^M
\frac{1}{|s_{i+1}-s_i|}\,,
\label{ffactor}
\ee
modulo a constant which does not depend on the $\phi_i$'s.
The appearance of this factor
 is due, shortly speaking,  to the specifics of
the Douglas minimization for stepwise functions $p(\theta(\phi))$
in contrast to smooth functions.
The points $\phi_i$'s, where the function has discontinuities, are
irregular points from the point of view of the minimization because
the minimizing function  $\theta_*(\phi)$ has to satisfy 
$\theta_*(\phi_i)=\phi_i$.
We can still perform a reparametrization $\phi\to\theta(\phi)$
at intermediate points 
$\phi \in (\phi_i,\phi_{i+1})$. For such a reparametrization 
$p(\theta(\phi))$ moves along the step but
its value remains unchanged. Therefore, this is a zero mode in
 Douglas' minimization and we have to integrate over these
 zero modes.  For smooth functions, $\theta_*(\phi)$ was just fixed.
The integration over the zero modes is exactly the same as the
integration over reparametrizations at intermediate points 
 described in Subsect.~\ref{ss:4.3} and Appendix~D. The result of
this integration at the interval $(\phi_i,\phi_{i+1})$ is 
given by the $i$-th multiplier in \eq{ffactor}, while the product runs over 
the labels of intervals.

The scattering amplitude then takes the form
\begin{eqnarray}
G(\Delta p_1,...,\Delta p_M)&\propto &
\prod_{i=1}^{M-1}\int\limits_{-\infty}^{\t_{i+1}}\frac{\d \t_i}{1+\t_i^2}
\prod_{i=1}^{M}\frac{1}{|\t_{i+1}-\t_i|}\,\left[
\frac{|\t_{i+1}-\t_{i}||\t_{i}-\t_{i-1}|}{|\t_{i+1}-\t_{i-1}|}\right]^{\Delta p_i^2/2\pi \K}
\non &&~~~~\times\exp \Big(\frac{1}{2\pi K}
\sum_{i,j=1 \atop i\neq j}^M\Delta p_i\Delta p_j\ln  |\t_i-\t_j|\Big)
\label{KN}
\end{eqnarray}
which looks similar to the Koba--Nielsen amplitude. It should be kept in mind 
that this expression is only valid if the area-law behavior of the Wilson loop 
dominates over other contributions to $W$, and if the number of external 
particles $M$ is large and/or $m$ is small compared to $\sqrt{\K}$ as is
already mentioned.

In Appendix~E we have discussed this expression in much more
details. In particular we have shown how the Veneziano amplitude with 
the usual Regge behavior follows.

\newsection{Large number of external particles WL/SA duality}

The reader may have noted that the area behaved Wilson loop and the
large $M$ amplitude before the integrations over the $\phi_i$'s
are done look very similar (see  Eqs.~(\ref{D7}) and (\ref{KN})), 
except that the
variables $z$ and $p$ are somehow interchanged. This reminds us about the
Wilson-loop/scattering-amplitude duality mentioned in the introduction,
which was found in SYM perturbation theory, see Refs.~\cite{AM07a} and
\cite{DSK07}.

This similarity can be made explicit by the substitution
\begin{equation}
z(\phi )=\frac{1}{K}~\sum_i p_i ~\Theta (\phi-\phi_i)\Theta (\phi_{i+1}-\phi),
\label{F1}
\end{equation}
where the $\Theta$'s are the Heaviside step functions. From this expression 
we have
\begin{equation}
\dot{z}(\phi)=\frac{1}{K}~\sum_i\Delta p_i~\delta (\phi-\phi_i )=
\frac{\dot{\theta}(\phi)}{K}~\sum_i \Delta p_i ~\delta (\theta (\phi)-
\theta_i),
\label{F2}
\end{equation}
where $\theta=\theta (\phi)$ is some reparametrization with $\theta (\phi_i)
=\theta_i$.
If this equation is inserted in the area behaved Wilson loop 
(\ref{D7}), it reproduces the multiparticle amplitude in (\ref{KN}) when
integrated over the $\phi_i$.  
The spatial variable on the left-hand side of Eq.~(\ref{F1})
is on the right-hand side composed of a number of constant vectors given by 
the momenta, which is
rather analogous to the duality (\ref{du}) mentioned in the
introduction for the ${\cal N}=4$ SYM.
A difference is that here we need to integrate over all trajectories
considered as functions of $\phi_i$.

Equation~(\ref{F1}) reemphasizes the often mentioned fact that for strings
the large $z$ limit is equivalent to the large $p$ limit. The 
surprising interchange of space and momentum is clearly a stringy effect.

It is easy to check that the expression (\ref{F1}) actually
satisfies Douglas' variational condition (\ref{=0}) or, alternatively,
Eq.~(\ref{==0}). To see this, let us make a variation
\begin{equation}
\phi(\theta)= \phi_* (\theta)+\delta\phi (\theta),
\end{equation}
with $\delta\phi$ small and $\phi_*$ giving the minimal Douglas functional, as
is explained in Appendix~B. We have the conditions
\begin{equation}
z(\theta_*(\phi_i))=x_i,
\label{F3}
\end{equation}
since the curve passes through the points $x_i$. Therefore in the variation we
need the conditions
\begin{equation}
\delta\phi(\theta_i)=0
\label{F4}
\end{equation}
for all $i$.
The first variational derivative of the Douglas functional is proportional to
\begin{equation}
\int\limits_0^{2\pi}\d\theta \ppint  \d\theta'\,
\delta\phi (\theta)\dot{z}(\theta)
\dot{z}(\theta')~\cot\frac{\phi_*(\theta)-\phi_*(\theta')}{2}.
\label{F5}
\end{equation}

Inserting Eq.~(\ref{F2}), the expression (\ref{F5}) becomes
\begin{equation}
\sum_{ij}\Delta p_i\Delta p_j~\delta\phi(\theta_i)~\cot\frac{\phi_i
-\phi_j}{2},
\label{F6}
\end{equation}
where we get contributions only from the fixed points $x_i$. 
Because of Eq.~(\ref{F4}) we see that the first variational derivative
(\ref{F6}) vanishes.
Hence the duality relation (\ref{F1}) actually satisfies Douglas' 
variational principle and in this sense the curves (\ref{F1}) represent
the dominant trajectories (the ``master trajectories'') in phase space when 
the momenta are given. The special trajectory (\ref{F1}) is a kind of zero
mode solution of the Douglas variational problem, since it does not actually
determine the function $\phi_*$, which is irrelevant in the case of
stepwise constant momenta. We have already discussed this specifics 
of step functions in Sect.~\ref{s:5}.

To sum up, the result is
\begin{eqnarray}
G(\Delta p_1,...,\Delta p_M)&\propto& \prod_{i=1}^{M-1}\int\limits_0^{\phi_{i+1}}
\d\phi_{i}\,\prod_{i=1}^M
\frac{\sin^2(\phi_i/2)}{\sin[(\phi_{i+1}-\phi_i)/2]} \, \non && ~~~~\times
W\Big(z(\phi )\rightarrow\frac{1}{\K} \sum_i p_i 
\Theta (\phi-\phi_i)\Theta (\phi_{i+1}-\phi)\Big),              
\end{eqnarray}
where $W(z(\phi))$ is the Wilson loop as a function of the boundary curve
$z(\phi)$.
Except for the integrations this is similar to the supersymmetric case 
(\ref{du}) discussed in \cite{AM07a} and \cite{DSK07}. 

This similarity can be made even more explicit by noting that Eq.~(\ref{F1})
means that in the interval $\phi_i <\phi <\phi_{i+1}$ the vector $z(\phi )$ is 
equal to the vector $p_i/\K$. Remembering that $z$ equals $x_i$ and
$x_{i+1}$ for the parameter values $\phi_i$ and  $\phi_{i+1}$, respectively.
This amounts to
\begin{equation}
\Delta p_i=\K\,(x_{i-1}-x_{i}),\qquad{\rm for}~~\phi_i<\phi<\phi_{i+1},
\end{equation}
in conformity with Eq.~(\ref{du}).

We again emphasize that all the above results are valid only
when the number of external particles $M$ is large. In the general case
of a smaller number of produced particles the situation is much more complex.
Presumably there will be important fluctuations around the master
trajectory giving contributions to the kernel $\cal K$.

\newsection{Conclusions}

We have found  a relation between the meson scattering 
amplitudes and the Wilson loop for large $N$ QCD. We then investigated the
behavior of an area behaved $W$
when a functional Fourier transform was performed, leading to the Veneziano 
multiparticle amplitude in the Koba--Nielsen formulation. This turned out to be
very useful when we inserted an area approximation for the Wilson loop in our 
general expression for the large $N$ QCD amplitude. The result is a convolution
integral, with the well-known Koba--Nielsen integrand convoluted with
a kernel $\cal K$. The usual poles always occur because they cannot be 
prevented by the kernel. 

For a very large number of external particles $\cal K$
becomes essentially a constant, and hence the Veneziano multiparticle
amplitude appears. It then turns out that there  exists a nice duality 
between Wilson loops and scattering amplitudes, somewhat similar to 
the supersymmetric case.

Although these phenomena are valid for large $N$, one might hope
that something similar occurs for $N=3$, in which case this would be
observable at LHC, where a huge number of particles are produced. So
we hope to see at least some tracks of the Veneziano amplitude
in the collider data!

\subsection*{Acknowledgments}

%%{\bf Acknowledgments}. 
We are indebted to Paolo Di~Vecchia, Alexander Gorsky, 
Alexey Kaidalov, Raffaele Marotta,  
Niels Obers, and Vladimir Zoller for useful discussions.

%%\eop
\setcounter{section}{0}

\appendix{An example of the use of the ``variable'' gamma matrix}

Here we shall give an example of how the gamma matrix $\gamma_\mu (t)$
operates in the case where we consider the quantity
\begin{equation}
F={\rm sp~P}\;\exp \left[\i\int_0^\tau ~\d t~ p_\mu (t) \gamma_\mu(t)\right]=
{\rm sp}~\prod_{j=n}^1\exp\left(\i {\not}p_{j}
(t_j-t_{j-1})\right),
\end{equation}
where we took the momenta $p$ to be stepwise constant. 
We can now use
\begin{equation}
\e^{\i\not p_j \Delta t_j}=\cos M\Delta t_j+\i\frac{\not p_j}{M}~\sin 
M\Delta t_j,~~~\Delta t_j=t_j-t_{j-1}.
\end{equation}
Here we took $p_j^2=-M^2$ for all $j$, thus assuming that all the external 
mesons have the same mass. Then
\begin{equation}
F={\rm sp}~\prod_{j=n}^1\left(\cos M\Delta t_j+\i\frac{\not p_j}{M}~\sin 
M\Delta t_j\right).
\end{equation}
For the two point case $n=2$ we get
\begin{equation}
F_{n=2}=4\Big(\cos M\Delta t_1\cos M\Delta t_2-\frac{p_1p_2}{M^2}~\sin M
\Delta t_1\sin M\Delta t_2\Big).
\end{equation}
Also
\begin{equation}
F_{n=3}=4\Big(c_1c_2c_3-\frac{p_1p_2}{M^2}s_1s_2c_1-\frac{p_1p_3}{M^2}s_1c_2s_3-
\frac{p_2p_3}{M^2}c_1s_2s_3\Big),
\end{equation}
and 
\begin{eqnarray}
F_{n=4}&=&4\Big(c_1c_2c_3c_4-\frac{p_3p_4}{M^2}c_1c_2s_3s_4-\frac{p_2p_4}{M^2}
c_1s_2
c_3s_4-\frac{p_3p_2}{M^2}c_1s_2s_3c_4
-\frac{p_1p_4}{M^2}s_1c_2c_3s_4\nonumber \\
&&-\frac{p_1p_3}{M^2}s_1c_2s_3c_4
-\frac{p_1p_2}{M^2}s_1s_2c_3c_4
+\frac{1}{M^2}
((p_1p_2)(p_3p_4)-(p_1p_3)(p_2p_4)\nonumber \\
&&+(p_1p_4)(p_2p_3))s_1s_2s_3s_4\Big).
\end{eqnarray}
Here $c_i=\cos M\Delta t_i$ and $s_i=\sin M\Delta t_i$.

If the momenta are not stepwise constant, we can use the above procedure
with $n\rightarrow\infty$, if we divide the interval from 0 to $\tau$ into 
intervals $\tau/n$ and take the limit at the end.

\appendix{Douglas' approach to the minimal area\label{appA}}

The Douglas algorithm~\cite{Dou31}
for finding the area of the minimal surface bounded by a 
closed contour $C$ which is parametrized by the function $x_\mu(\sigma)$ 
is based on minimizing the boundary functional%
\footnote{For a modern review see also Ref.~\cite{Mig93}, Appendix~H.}
\begin{equation}
A[x(\theta)]=\frac{1}{8\pi}\int\limits_ 0^{2\pi} \d \sigma 
\int\limits_ 0^{2\pi} \d \sigma^\prime \, 
\frac{\left[x(\theta(\sigma))-x(\theta(\sigma^\prime))\right]^2}
{1-\cos(\sigma-\sigma^\prime)} 
\label{D}
\end{equation}
with respect to the reparametrizing functions $\theta(\sigma)$ 
($\d \theta(\sigma)/\d \sigma\geq 0$). 
The numeric value of $A$ for the given $C$
depends on the choice of $\theta(\sigma)$ and in general 
\begin{equation}
A[x(\theta)]\geq S_{\rm min}(C)
\end{equation}
while the equality is reached for certain function 
$\theta(\sigma)=\theta_*(\sigma)$ which provides the minimum of $A$.
The function $\theta_*(\sigma)$ is of course contour-dependent. 

To prove the fact that
\begin{equation}
A[x(\theta_*)]= S_{\rm min}(C)\,,
\label{A=S}
\end{equation}
we reconstruct $X_\mu(r,\sigma)$, describing the surface, in the interior
of the unit disk $r<1$ from the boundary value 
$X_\mu(1,\sigma)=x_\mu\left(\theta_*(\sigma)\right)$ by the Poisson formula and
note that thus constructed $X_\mu$ automatically obeys conformal gauge
\begin{equation}
\frac{\partial X}{\partial r}\cdot \frac{\partial X}{\partial \sigma}=0\,,
\qquad 
r^2\frac{\partial X}{\partial r}\cdot \frac{\partial X}{\partial r}=
\frac{\partial X}{\partial \sigma}\cdot 
\frac{\partial X}{\partial \sigma}\,.
\label{conformal}
\end{equation}
Therefore, the Nambu--Goto action coincides for this configuration with
the quadratic action and the boundary action in Polyakov string theory 
coincides with the area.

By varying $A[\theta]$ with respect to $\theta(\sigma)$ at given $C$,
we get the following equation for $\theta_*(\sigma)$:
\begin{equation}
\ppint \d\sigma^\prime \,
\frac
{\dot x(\theta_*(\sigma))\cdot
\left[x(\theta_*(\sigma))-x(\theta_*(\sigma^\prime))\right]}
{1-\cos(\sigma-\sigma^\prime)}=0 
\label{=0}
\end{equation}
which can be written in several equivalent forms.

To illustrate how \eq{=0} can be used to determine  $\theta_*(\sigma)$,
let us consider the case of a plane contour, when the problem can be 
solved by a conformal map, and concentrate on the case of an ellipse
\be
x_1=a \cos \theta(\sigma)\,,\qquad x_2=b \sin \theta(\sigma)\,.
\label{ellipse}
\ee
Then \eq{=0} takes the form
\begin{equation}
\ppint \d\alpha\, 
\frac{\sin\left[\theta_*(\sigma+\alpha)-\theta_*(\sigma)\right]}{1-\cos\alpha}=
\epsilon \ppint \d\alpha\, 
\frac{\sin\left[\theta_*(\sigma+\alpha)+\theta_*(\sigma)\right]
-\sin\left[2\theta_*(\sigma)\right]}{1-\cos\alpha}
\label{ellipseeq}
\end{equation}
with 
\be
\epsilon=\frac{a^2-b^2}{a^2+b^2}\,.
\ee

For the simplest case of a circle ($a=b$), the coordinates $r, \sigma$
are conformal so that
\be
\theta_*(\sigma)=\sigma \qquad \fbox{circle} \,.
\label{circlephi}
\ee
The right-hand side of \eq{ellipseeq} vanishes for $\epsilon=0$, while
the left-hand side also vanishes for $\theta_*(\sigma)$ given by 
\eq{circlephi}.  

For $\epsilon\neq 0$ the following ansatz passes through \eq{ellipseeq}:
\be
\theta_*(\sigma)=\sigma +\sum_{n\geq1} c_n \sin (2n \sigma)
\qquad \fbox{ellipse} \,,
\label{ellipsephi}
\ee
after which a set of algebraic equations relating $c_n$'s emerges.
For small deforming parameter $\epsilon$, its iterative solution to order
${\cal O}(\epsilon^7)$ found by Mathematica is
\bea 
c_1 &=& \epsilon - \epsilon^3/4 + \epsilon^5/8+{\cal O}(\epsilon^7)\non
c_2 &= &3\epsilon^2/4-5 \epsilon^4/8 + 25 \epsilon^6/64+{\cal O}
(\epsilon^7)\non
c_3 &= &5\epsilon^3/6 - 5 \epsilon^5/4+{\cal O}(\epsilon^7)\non
c_4 &= &35\epsilon^4/32 - 77 \epsilon^6/32+{\cal O}(\epsilon^7)\non
c_5 &= &63\epsilon^5/40+{\cal O}(\epsilon^7)\non
c_6 &= &77\epsilon^6/32+{\cal O}(\epsilon^7)\,.
\label{itera}
\eea
The minimal area 
\be
S_{\rm min}=\pi a b=\pi a^2 \sqrt{\frac{1-\epsilon}{1+\epsilon}}
\ee 
is of course reproduced to this order
by substituting~\rf{ellipsephi}, \rf{itera} into \eq{D}.

Integrating by parts,
we can also rewrite the Douglas functional~\rf{D} as 
\be
A=-\frac 1{4\pi} \int\limits_0^{2\pi}\d\theta_1 
\int\limits_0^{2\pi} \d\theta_2\,
\dot x(\theta_1)\cdot \dot x(\theta_2)\,\ln 
\left(1-\cos\left[\sigma(\theta_1)-\sigma( \theta_2)\right]  \right).
\ee
Its variation with respect to $\sigma(\theta)$ results in the 
equation
\be
\ppint \d\alpha \;
\dot x(\theta)\cdot \dot x(\alpha)\,\cot 
\left(\frac{\sigma_*(\theta)- \sigma_*(\alpha)}2  \right)=0
\label{==0}
\ee
which is nothing but the Douglas original equation from the 
Abstract of 1927.

By solving this equation, we obtain the function $\sigma_*(\theta)$ 
which is inverse to $\theta_*(\sigma)$, given by \eq{ellipsephi}, and reads
\be
\sigma_*(\theta)=\theta +\sum_{n\geq1} d_n \sin (2n\theta)\qquad
\fbox{ellipse}
\ee
with
\bea
d_1 &=& -\epsilon +{\cal O}({\epsilon^7})\non
d_2 &=& \epsilon^2/4 + \epsilon^4 /8 + \epsilon^6/16+{\cal O}({\epsilon^7})\non
d_3 &=& -\epsilon^3/12- \epsilon^5/16+{\cal O}({\epsilon^7}) \non
d_4 &=& \epsilon^4/32+ \epsilon^6 /32+{\cal O}({\epsilon^7})\non
d_5 &=& -\epsilon^5/80+{\cal O}({\epsilon^7})\non
d_6 &=& \epsilon^6/192+{\cal O}({\epsilon^7}) \,.
\eea

To proceed further, 
it is convenient to use a standard ellipse of the area $\pi$ with
\be
a=\sqrt[4]{\frac{1+\epsilon}{1-\epsilon}}\,,\qquad
b=\sqrt[4]{\frac{1-\epsilon}{1+\epsilon}}\,.
\ee
Then
\bea
x_1(\sigma)= a \cos \theta(\sigma)&=& \sum_{n\geq1} \mu_n \cos (2n-1)\sigma \non
x_2(\sigma)= b \sin \theta(\sigma)&=& \sum_{n\geq1} \nu_n \sin (2n-1)\sigma 
\label{x1x2}
\eea
and from \eq{itera} $\mu_n=\nu_n$ to order ${\cal O}(\epsilon^7)$.

This can be understood if we continue 
the boundary coordinates~\rf{x1x2} inside the unit circle as
\bea
X_1(r,\sigma)&=& \sum_{n\geq1} \mu_n \, r^{2n-1}\cos (2n-1)\sigma \non
X_2(r,\sigma)&=& \sum_{n\geq1} \nu_n \, r^{2n-1}\sin (2n-1)\sigma \,.
\eea
It can be then explicitly verified that these coordinates obeys the conformal
gauge~\rf{conformal} for $\mu_n=\nu_n$.

Introducing the analytic function
\be
{\cal M}(z)=\sum_{n\geq1} \mu_n z^{2n-1}\,, 
\label{calM}
\ee
%with positive odd $\mu_n$'s~\cite[P-J]
that describes a conformal map of a unit disk onto the interior of 
the standard ellipse, we get the following algebraic equation
\be 
a \cos \theta(\sigma) + \i b \sin \theta(\sigma)=
{\cal M}\left(\e^{\i \sigma}\right)
\label{ellipsemap}
\ee
or
\be
\e^{\i \theta(\sigma)}=\frac{{\cal M}\left(\e^{\i \sigma}\right)+
\sqrt{{\cal M}^2\left(\e^{\i \sigma}\right)-(a^2-b^2)}}{a+b}
\label{eq1}
\ee
which determines $\theta(\sigma)$ for the given ${\cal M}(z)$.
For the ellipse, $\theta$ plays the role of an angular
variable in the parametrization~\rf{ellipse}, while the function
$\sigma_*(\theta)$ relates it to the variable $\sigma$ inherited from
the conformal coordinates $r$, $\sigma$ obeying \eq{conformal}.

An analytic function that describes the conformal map of a unit disk onto 
the interior of an ellipse was found by Schwarz in 1869~\cite{boo} and gives
\be
{\cal M}(z)= \sqrt{a^2-b^2} \sin \left[ 
\frac{\pi}{2K(s)} F\left(\frac{z}{\sqrt{s}};s \right) 
\right],
\label{Schwarz}
\ee
where 
\be
F\left(z;s \right) = \int\limits_0^z \frac{\d x}{\sqrt{(1-x^2)(1-s^2 x^2)}}
\ee
and
\be
K(s)\equiv F\left(1;s\right)=
\int\limits_0^1 \frac{\d x}{\sqrt{(1-x^2)(1-s^2 x^2)}}
\ee
are, respectively, the incomplete and complete elliptic integrals of
the first kind.%
\footnote{We use the notations for elliptic integrals from Wikipedia. 
They are related to those of Mathematica as 
$F\left( z;s \right)={\rm EllipticF}[{\rm ArcSin}[z],s^2]$
and $K(s)={\rm EllipticK}[s^2]$.} 
The parameter $s$ is related to $\epsilon$ by
\be
\log \frac{a+b}{a-b}= 2\xi(s)\equiv
\frac{\pi K\left(\sqrt{1-s^2}\right)}{2K\left(s\right) } 
\label{Gr''otzsch}
\ee
so that iteratively
\be
s=2\epsilon -\frac 32 \epsilon^3 +\frac 12 \epsilon^5 -\frac1{32}\epsilon^7
+\frac {3}{128} \epsilon^9+{\cal O}(\epsilon^{11})\,.
\label{s2e}
\ee

Substituting the function~\rf{Schwarz} into \eq{ellipsemap}, we obtain
the equation
\be
\frac{\pi}2 -\theta +\i \xi = \frac{\pi}{2K(s)}
 F\left(\frac{\e^{\i\sigma}}{\sqrt{s}};s \right), 
\label{neweq}
\ee
whose imaginary part reproduces \eq{Gr''otzsch} in view of the important 
identity
\be
F\left(\frac{\e^{\i\sigma}}{\sqrt{s}};s \right) = 
F\left(\frac{\e^{-\i\sigma}}{\sqrt{s}};s \right)+\i K\left(\sqrt{1-s^2}\right).
\ee
The real part of \eq{neweq} determines the function $\theta(\sigma)$.

In order to obtain it, it is convenient first to rewrite \eq{Schwarz} as 
\be
{\cal M}(z)= \sqrt{a^2-b^2} \cosh \left[ 
\frac{\pi}{2K(s)} \int\limits_0^{{\rm arccosh}(z/\sqrt{s})}
\frac{\d \lambda}{\sqrt{1-s^2 \cosh^2 \lambda}}\right].
\label{Schwarz1}
\ee
This form is more suitable for the case when the first argument of $F$ 
in \eq{Schwarz} is large as it is for small $\epsilon$ and correspondingly
for small $s$, related by \eq{s2e}.
Substituting the function~\rf{Schwarz1} into \eq{ellipsemap}, we obtain
the equation
\be 
\xi+\i \theta=\frac{\pi}{2K(s)}  
\int\limits_0^{{\rm arccosh}(\e^{\i \sigma}/\sqrt{s})}
\frac{\d \lambda}{\sqrt{1-s^2 \cosh^2 \lambda}}  
\label{A.29}
\ee
which is equivalent to \eq{neweq}.

Differentiating~\eq{A.29}, we obtain
\be
\theta^\prime(\sigma)=\frac{\pi}{2K(s)} \frac{1}{\sqrt{1+s^2 -2s\cos 2 \sigma}} 
=\frac{\pi}{2K(s)} \frac{1}{\sqrt{(1-s)^2 +4 s\sin^2 \sigma}} 
\label{deriva}
\ee
which yields 
\be
\theta(\sigma) = 
%%\frac{\pi}{2K(s)} \int\limits_0^{\sigma}
%%\frac{\d \sigma'}{\sqrt{1+s^2 -2 s\cos 2 \sigma'}} = 
\frac{\pi}{2K(s)} 
\int\limits_0^{\sigma}
\frac{\d \sigma'}{\sqrt{(1-s)^2 +4 s\sin^2 \sigma'}} 
=\frac{\pi}{2K(s)} \frac1{(1-s)}
F\left(\sin \sigma\,; \frac{2\i \sqrt{s}}{(1-s)} \right)
\label{minfun}
\ee
which is again an elliptic integral of the first kind.
This reproduces iteratively \eq{itera} provided \eq{s2e} is satisfied.

The solution \rf{minfun} obeys the properties required for the Douglas
minimizing function $\theta_*(\sigma)$. It is seen from \eq{deriva} that
$\theta^\prime_*(\sigma)$ is positive and finite for $s<1$, as is required
for a reparametrization. For $s\to0$ we have $\theta^\prime_*(\sigma)\to 1$
as it should for a circle.
In the limit $s\to1$ when $K(1)=\infty$
we have $b\to0$, so the ellipse collapses. Then $\theta^\prime_*(\sigma)$ 
vanishes everywhere except for $\theta=0,\pi$ where it becomes
infinite:
\be
\theta^\prime_*(\sigma)\stackrel{b\to0}\to
\pi \delta(\sigma) +\pi \delta (\sigma-\pi)\,,
\ee
while
\be
\theta_*(\sigma)\stackrel{b\to0}\to\pi \Theta(\sigma) +\pi \Theta (\sigma-\pi)
\ee
is stepwise.

\appendix{The pre-exponential in \eq{circle1}}

To calculate the pre-exponential in \eq{circle1},
we substitute
\be
\theta(\sigma)=\theta_*(\sigma)+\beta(\sigma)
\ee
or
\be
\sigma(\theta)=\sigma_*(\theta)+\beta(\theta) \,,
\ee
where
\be
\beta(0)=\beta(2\pi)=0
\label{bbc}
\ee
and expand the exponent in \eq{diskxdisktheta} (or \eq{diskxdisktheta1}) to
the quadratic order in $\beta$. The linear in $\beta$ term vanishes 
because $\theta_*(\sigma)$ is the extremum while the quadratic part
reads
\be
S_2[\beta(\theta)]= \frac{\K}{2} \int \d \theta_1 \d \theta_2 \,
\dot {\vec x}(\theta_1) \dot{\vec x}(\theta_2)\,
G^{-1}\left(\sigma_*(\theta_1)-\sigma_*(\theta_2)\right) 
\left[\beta(\theta_1)\beta(\theta_2)-\beta^2(\theta_1)\right].
\label{S2}
\ee
The function $\beta(\theta)$ has to obey
\be
\beta^\prime (\theta) \geq -\sigma_*^\prime(\theta)
\label{restriction}
\ee
for the derivative of the reparametrizing function to be positive.
This is always satisfied if $\beta$ is small and smooth enough.
Therefore, the measure for the Gaussian path integration over $\beta(\theta)$
is the usual one for smooth functions $\beta(\theta)$ but 
we shall see subtleties for the functions with large derivative.

In order to calculate the pre-exponential in \eq{circle1}, we need
to do the Gaussian integral
\be
I_2= \int \D\beta(\theta) \e^{-S_2[\beta(\theta)]}
\label{I2}
\ee
with $S_2[\beta(\theta)]$ given by \eq{S2}.

For a circle of the radius $R$, when
\be
x_1(\theta)= R \cos \theta \,,\quad x_2(\theta)= R \sin \theta \,,\quad 
x_3(\theta)=x_4(\theta)=0
\ee
and $\sigma_*(\theta)=\theta$ according to \eq{circlephi},
we have
\be
S_2[\beta(\theta)]= -\frac{\K R^2}{4\pi} 
\int_0^{2\pi}\d \theta_1 \d \theta_2 \,
\frac{\cos\left(\theta_1-\theta_2\right)}
{1-\cos\left(\theta_1-\theta_2\right)} 
\left[\beta(\theta_1)\beta(\theta_2)-\beta^2(\theta_1)\right].
\label{S2R}
\ee

It is now possible to calculate the path integral over $\beta(\theta)$ by
the mode expansion
\be
\beta(\theta)=a_0+ \sum_{n=1}^\infty 
\left(a_n \cos n\theta + b_n \sin n\theta \right),
\label{betaexp}
\ee
where $a_0=-\sum_{i=1}^\infty a_n$ to obey the boundary condition~\rf{bbc}
and $a_n$, $b_n$ have to satisfy
\be
\sum_{n=1}^\infty n (- a_n \sin n\theta +b_n \cos n \theta ) \geq -1  
\label{rest1}
\ee
for the restriction \rf{restriction} to be fulfilled.
Inserting \eq{betaexp} into \eq{S2R}, we find
\be
S_2= \frac{\pi\K R^2}{2} \sum_{n=1}^\infty (n-1) \left(a_n^2 +b_n^2\right).
\label{S2ab}
\ee

A consequence of \eq{S2ab} is that $a_1$ and $b_1$ are zero modes.
They are, however, restricted by \rf{rest1} as $-1<a_1,b_1<1$, so
we believe that the integrals over $a_1$ and $b_1$ simply give a constant.
The integrals over nonzero modes are Gaussian and if we were not
take into account the restriction~\rf{rest1}, the result would be
\be
I_2\propto \prod_{n=2}^\infty \int \d a_n \d b_n\,\e^{-S_2}=
\prod_{n=2}^\infty \left[ \K R^2(n-1)/2\right]^{-1}\,.
\label{I2prod}
\ee

The infinite product in \eq{I2prod} can be calculated in the standard way, 
using a regularization via the $\zeta$-function:
\bea
\prod_{n=1}^\infty A &=&A^{\zeta(0)}=A^{-1/2}\,, \non
\prod_{n=1}^\infty n &=& \sqrt{2\pi}\,.
\label{zetareg}
\eea
We then obtain from \eq{I2prod} 
\be
I_2\propto (\K R^2)^{1/2}
\label{I2fin}
\ee
which fixes the $\K R^2$ dependence of the pre-exponential in \eq{circle1}.

Since the typical values of $\beta$, which
are essential in the path integral over $\beta$ in \eq{I2},
are $\beta\sim 1/\sqrt{K} R$, i.e. 
 small for $\sqrt{K} R >>1$, the higher terms of an expansion 
of $A[\theta_*(\sigma)+\beta]$ in
$\beta$ are suppressed at large $\sqrt{K} R$.
The loop expansion goes in the parameter $1/KR^2$ and only one loop
contributes with the given accuracy. We thus reproduce
the behavior of the type in \eq{circle1}.

We can now ask the question whether or not the typical values of $a_n$'s and 
$b_n$'s, which are essential in the integral over the nonzero modes in
\eq{I2prod}, obey the restriction \rf{rest1}. We estimate these values as
\be
a_n\sim b_n \sim (n \K R^2)^{-1/2}
\ee
which are small in accordance with the standard wisdom that high modes
are not essential in the Gaussian path integral. 
However, the restriction \rf{rest1} can be satisfied by a single mode only if 
\be
n \la \K R^2 \,. 
\ee
We assume that this type of the restriction on the number of modes
becomes unimportant for $\K R^2\to \infty$. 

\appendix{Path integrals over $Diff(\mathbb{R})$ }
%%%%%%{Properties of the measure on $Diff(\mathbb{R})$ }

The measure for integrating over reparametrizations is determined 
by the metric~\cite{Pol87}
\begin{equation}
\| \delta t \|^2 = \int\limits_{-\infty}^{+\infty} \d s\,
t'(s) \left[\delta t(s)\right]^2 
\label{metric}
\end{equation}
which is invariant under reparametrizations.
It differs from the usual one
\begin{equation}
\| \delta t \|^2 = \int\limits_{-\infty}^{+\infty} \d s\,
\left[\delta t(s)\right]^2 
\label{metric1}
\end{equation}
by the presence of $t'(s)=\d t(s)/\d s$ in the integrand. 

An explicit representation of the measure 
$\Dp t(s)$ for the integration over 
reparametrizations is given by \eq{o.i.t}.
As distinct from the usual measure $\D t(s)$:
\be
\int \D t(s) \cdots
= \lim_{L\to\infty}\int\limits_{-\infty}^{+\infty} 
\prod_{j=1}^{L}{\d t_j} \cdots \,,
\label{to.i.usual}
\ee
the integrals in \eq{o.i.t} are ordered.

The measure \rf{o.i.} is the invariant measure on the group
$Diff(S^1)$ of reparametrizations (diffeomorphisms) of a circle.
Analogously the measure \rf{o.i.t} is the invariant measure on the group
$Diff(\mathbb{R})$ of reparametrizations (diffeomorphisms) of the real
axis. They are defined in the way to be invariant under the %%(infinitesimal)
$PSL(2;\mathbb{R})$ projective transformation at very small but {\em finite}\/ 
discretization spacings $\varepsilon_i=s_i-s_{i-1}\sim 1/L$. 
This guarantees the invariance of the measure under reparametrizations
in the limit $L\to\infty$.

If the integrand of the path integral over reparametrizations is not a 
functional but a function of, say, only $t(s_j)= t_j$ and $t(s_k)= t_k$ 
with $j>k$, we can integrate over intermediate values $t_i$'s with
$k<i<j$. Analogously, for a function of $M$ variables we integrate
over $t_i$'s inside $M$ intervals. The result of such an integration
is a function of the remaining $M$ variables $t_{i_1}$, \ldots,
$t_{i_M}$, which should be covariant under the projective
transformation. 
%%We expect that it may be possible to explain the appearance of 
%%the measure of the type in \eq{measure} along this line.

The simplest integral we meet is of the type
\be
\int\limits_{t_{i-1}}^{t_{i+1}} \d t_i\,\frac{1}{(t_{i+1}-t_i)(t_{i}-t_{i-1})} 
\ee
which is logarithmically divergent at the upper and lower limits of
the integration. We regularize it by introducing a small $\delta $ as
\be
\lim_{\delta\to0}\int\limits_{t_{i-1}}^{t_{i+1}}\d t_i\,  
\frac{\delta}{(t_{i+1}-t_i)^{1-\delta}(t_{i}-t_{i-1})^{1-\delta}} 
=\frac{2}{(t_{i+1}-t_{i-1})}\,. 
\ee

We can  now perform a heuristic derivation of the measure 
\rf{measure11}, which is 
%%similar to the measure~\rf{measure} and is also
invariant under the projective transformations, using
the following formula
%%\be
%%\int\limits_{t_0}^{t_K} \d t_i\, 
%%\frac{[\Gamma(\varepsilon (K-i))]^{-1}}{(t_K-t_i)^{1-(K-i)\varepsilon }}
%%\frac{[\Gamma(\varepsilon i)]^{-1}}{(t_i-t_0)^{1-i\varepsilon }} =
%% \frac {[\Gamma(\varepsilon K)]^{-1}}{(t_i-t_0)^{1-K\varepsilon }} \,,
%%\qquad K>i
%%\label{rema}
%%\ee
%%that does the job if $K\varepsilon \ll 1$.
\be
\int\limits_{t_0}^{t_K} \d t_i\, 
\frac{[\Gamma(\nu_1)]^{-1}}{(t_K-t_i)^{1-\nu_1 }}
\frac{[\Gamma(\nu_2)]^{-1}}{(t_i-t_0)^{1-\nu_2 }} =
 \frac {[\Gamma(\nu_1+\nu_2)]^{-1}}
{(t_K-t_0)^{1-\nu_1-\nu_2}} \,.
\label{rema}
\ee
Equation~\rf{rema} is an analogue of the well-known formula
\be
\int\limits_{-\infty}^{+\infty} \frac{\d t_i }{\sqrt{2\pi}}
\frac{\e^{-(t_K-t_i)^2/2\nu_1}}{\sqrt{\nu_1}}
\frac{\e^{-(t_i-t_0)^2/2\nu_2}}{\sqrt{\nu_2}}
=\frac{\e^{-(t_K-t_0)^2/2(\nu_1+\nu_2)}}
{\sqrt{(\nu_1+\nu_2)}}
\ee
which is used for calculations of path integrals with the usual 
Wiener measure.

Choosing in \eq{rema} $\nu_1= \varepsilon_1\delta$ and 
$\nu_2=\varepsilon_2\delta $ and repeatedly integrating over intermediate
points $t_1$, \ldots, $t_{K-1}$, we obtain
\be
\lim_{\delta\to0}\prod_{i=1}^{K-1} %% \left[
\int\limits_{t_{0}}^{t_{i+1}}\d t_{i}\,  
\frac{\varepsilon_{i+1}\delta }{(t_{i+1}-t_{i})^{1-\varepsilon_{i+1}\delta}}
%%\right]
\frac{\varepsilon_{1}}{(t_{1}-t_{0})^{1-\varepsilon_{1}\delta}} 
=\frac{\sum_{i=j}^M \varepsilon_j}{(t_{K}-t_{0})}=
\frac{(s_K-s_0)}{(t_{K}-t_{0})}\,. 
\label{oof}
\ee
Equation~\rf{measure11} can be now derived fixing
certain $M$ values of $t_i$'s and repeatedly integrating over  
 the intermediate points.

In order to justify \eq{simple}, we again introduce $L$ infinitesimal
intervals $\varepsilon_i$ and consider a step function $p(t(s))$ which has
discontinuities only at {\em finite}\/ number $M$ of points $t_{K_j}$ 
($j=1,\ldots,M$), so $\Delta p_i=0$ at all other points which we call the
intermediate points. To emulate   
the principal-value integral in the exponent in \eq{diskp}, we omit the
terms when $s_1$ and $s_2$ lie at two adjacent infinitesimal intervals.
Similarly to \eq{piip} we get 
\begin{eqnarray}
\lefteqn{ %%%\int\limits_{-\infty}^{+\infty} \d s_1 \pinti  \d s_2 \, 
%%%\frac{\left[ p(t(s_1))-p(t(s_2))\right]^2}{(s_1-s_2)^2} =
\sum_{k,j=1 \atop k\neq l\pm1}^L\; \int\limits_{t_{k}}^{t_{k+1}} \d \t_1
\int\limits_{t_{l}}^{t_{l+1}} \d \t_2\, \frac{(p_k-p_l)^2}{(\t_1-\t_2)^2} } \non &= &
\lim_{L\to\infty} \left[-2 \sum_{j,l=1 \atop j\neq l}^L \Delta p_{j}\cdot \Delta p_{l}
\log {|t_j-t_l|} -2 \sum_{j=1}^L  \Delta p_{j}^2 \log 
\frac{(t_{j+1}-t_{j-1})(t_j-t_{j-1})}{(t_{j+1}-t_{j-1})}
\right]  \nonumber \\
 &=&-2\left[ \sum_{j,l=1 \atop j\neq l}^M \Delta p_{K_{j}}\cdot \Delta p_{K_{l}}
\log {|t_{K_j}-t_{K_l}|} + \sum_{j=1}^M  \Delta p_{K_{j}}^2 \log 
\frac{(t_{K_{j+1}}-t_{K_{j-1}})(t_{K_j}-t_{K_{j-1}})}{(t_{K_{j+1}}-t_{K_{j-1}})}
\right],  \non &&
\label{uf}
\end{eqnarray}
where we have substituted $t_j$'s for $s_j$'s and
taken into account that $ \Delta p_{i}=0 $ for the intermediate 
points with $i\neq {K_{j}}$. 

We still have to insert \rf{uf} in the exponential and to integrate over
the intermediate points. Let us consider a piece from the point
$K_1$ to the point $K_3$ given by the ordered integral of the type
\bea
&& \hspace*{-5mm}\lim_{\delta\to0}
\int\limits_{t_i<t_{i+1}}
 \frac{\varepsilon_{K_1}}{|t_{K_1}-t_{K_1+1}|^{1-\varepsilon_{K_1}\delta}}
\prod_{i=K_1+1}^{K_2-1} \d t_i\, 
\frac{\varepsilon_i \delta}{|t_{i}-t_{i+1}|^{1-\varepsilon_i \delta}}
\left[\frac{|t_{K_2-1}-t_{K_2}||t_{K_2}-t_{K_2+1}|}
{|t_{K_2-1}-t_{K_2+1}|}\right]^{\alpha'\Delta p_{K_2}^2} \non
&&\hspace*{3cm} \times 
\frac{\varepsilon_{K_2}}{|t_{K_2}-t_{K_2+1}|^{1-\varepsilon_{K_2} \delta}}
\prod_{i=K_2+1}^{K_3-1} \d t_i\,  
\frac{\varepsilon_i \delta}{|t_{i}-t_{i+1}|^{1-\varepsilon_i \delta}}\,.
\label{uuf}
\eea
The integrals over all intermediate points, except for the ones with 
$i=K_2-1$ and $i=K_2+1$, are the same as in \eq{oof} and are
easily doable. The two remaining integrals are
\begin{eqnarray}
{{\rm (\ref{uuf})}}&= &\lim_{\delta\to0}
\int\limits_{t_{K_1}}^{t_{K_2}} \d t_{K_2-1} \int\limits_{t_{K_2}}^{t_{K_3}} \d t_{K_2+1}
\frac{{\cal E}_{1}}{|t_{K_1}-t_{K_2-1}|^{1-{\cal E}_1\delta}}
 \frac{\varepsilon_{K_2-1}\delta}{|t_{K_2-1}-t_{K_2}|^{1-\varepsilon_{K_2-1}\delta}} \non
&&\times \left[\frac{|t_{K_2-1}-t_{K_2}||t_{K_2}-t_{K_2+1}|}
{|t_{K_2-1}-t_{K_2+1}|}\right]^{\alpha'\Delta p_{K_2}^2}
\frac{\varepsilon_{K_2}\delta}{|t_{K_2}-t_{K_2+1}|^{1-\varepsilon_{K_2}\delta}}
\frac{{\cal E}_{2}}{|t_{K_2+1}-t_{K_3}|^{1-{\cal E}_1\delta}}
 \,, \non &&
\label{uuuf}
\end{eqnarray}
where
\be
{\cal E}_1=\sum_{i=K_1}^{K_2-2} \varepsilon_i\,,\qquad
{\cal E}_2=\sum_{i=K_2+1}^{K_3-1} \varepsilon_i \,.
\ee

For nonvanishing $\Delta p_{K_2}^2$ the integrals in  \eq{uuuf} 
differ from the one in \eq{oof}. 
However, they are also easily calculable for
$\delta\to0$ when only the domains $(t_{K_3}-t_{K_2+1})\to0$ and 
$(t_{K_2-1}-t_{K_1})\to0$ contribute. We get finally
\be
{{\rm (\ref{uuuf})}}
\propto\frac1{|t_{K_1}-t_{K_2}|} 
\left[\frac{|t_{K_1}-t_{K_2}|
|t_{K_2}-t_{K_3}|}
{|t_{K_1}-t_{K_3}|} \right]^{\alpha'\Delta p_{K_2}^2}
\frac1{|t_{K_2}-t_{K_3}|}\,,
\end{equation}
%\be
%{{\rm (\ref{uuuf})}}=
%\propto \left[\frac{|t_{K_1}-t_{K_2}|^{\alpha'\Delta p_{K_2}^2}
%|t_{K_2}-t_{K_3}|^{\alpha'\Delta p_{K_2}^2}}
%{|t_{K_1}-t_{K_3}|^{\alpha'\Delta p_{K_2}^2}} \right]\,.
%\end{equation}
thus proving \eq{simple}.

\appendix{Regge behavior of QCD scattering amplitude~\rf{KN}}
%%{Regge behavior without $PSL(2;\mathbb{R})$ symmetry}
We show in this appendix that the scattering amplitude (\ref{KN}) 
has asymptotic Regge behavior and reduces to the Veneziano amplitude for $M=4$.

The integrand in \eq{KN} is the same as in \eq{simple} except for the 
additional factors $1/(1+s_i^2)$ in the measure.
For the 4-point function we obtain 
\begin{eqnarray}
{G_4} 
&=&\left[~
\int\limits_{-\infty}^{\t_4} \d \t_3\int\limits_{-\infty}^{\t_3} \d \t_2
\int\limits_{-\infty}^{\t_2} \d \t_1
+\int\limits_{-\infty}^{\t_4} \d \t_3\int\limits^{\t_3}_{-\infty} \d \t_2
\int\limits^{+\infty}_{\t_4} \d \t_1 \right.\non &&\left.~~~~
+\int\limits_{-\infty}^{\t_4} \d \t_3\int\limits^{\infty}_{\t_4} \d \t_2
\int\limits^{\t_2}_{\t_4} \d \t_1
+\int\limits^{+\infty}_{\t_4} \d \t_3\int\limits_{\t_4}^{\t_3} \d \t_2
\int\limits_{\t_4}^{\t_2} \d \t_1\right]  
\nonumber \\* &&~~\times\frac{1}{(1+\t_1^2)(1+\t_2^2)(1+\t_3^2)} 
\frac1{|\t_{43}||\t_{32}||\t_{21}||\t_{41}|}  
\left( \frac{\t_{21} \t_{43}}{\t_{31} \t_{42}} \right)^{-\alpha' s}
\left( \frac{\t_{41} \t_{32}}{\t_{31} \t_{42}} \right)^{-\alpha' t}\!\!\!,
\non &&
\label{a40}
\end{eqnarray}
where we have introduced the notation
\be
\t_{ij}=\t_i-\t_j.
\ee
In \eq{a40} the integration is only over $\t_1$, $\t_2$ and $\t_3$, while
$\t_4$ is chosen in a way to preserve the cyclic symmetry but otherwise 
arbitrary.
The difficulty in analyzing the amplitude~\rf{a40} is the
presence of the factors $1/(1+\t_i^2)$ which violate the 
$PSL(2;\mathbb{R}$) symmetry. 

Let us introduce the variable
\be
x= \frac{\t_{21} \t_{43}}{\t_{31} \t_{42}}
\ee 
which runs from $0$ at $\t_2=\t_1$ to $1$ at $\t_2=\t_3$ and vise versa.
We can use $x$ instead of $\t_2$:
\begin{equation}
\t_2=\t_4-\frac{\t_{41}\t_{43}}{\t_{43}+x \t_{31}}
\end{equation}
and
\begin{equation}
\frac{\d \t_2}{(1+\t_2^2)} =\frac{\t_{43}\t_{31}\t_{41} \, \d x } 
%%{(\t_{43}+x \t_{31})^2}
{\t_{43}^2(1+\t_1^2)+2x\t_{43}\t_{31}(1+\t_1\t_4)+x^2 \t_{13}^2(1+\t_4^2)}\,.
\end{equation}
For the amplitude~\rf{a40} we get
\begin{eqnarray}
\lefteqn{
G_4=\int\limits_0^1 \d x\,  x^{-\alpha' s-1}(1-x)^{-\alpha' t-1} \left[
\int\limits_{-\infty}^{\t_4} \d \t_3\int\limits_{-\infty}^{\t_3} \d \t_1
+\int\limits_{-\infty}^{\t_4} \d \t_3\int\limits^{+\infty}_{\t_4} \d \t_1
+\int\limits^{+\infty}_{\t_4} \d \t_3\int\limits_{\t_4}^{\t_3} \d \t_1\right]}  
\nonumber \\* &&~\times\frac{1}{(1+\t_1^2)(1+\t_3^2)} 
\frac1{|\t_{43}||\t_{31}||\t_{41}|} \frac{(\t_{43}+x \t_{31})^2} 
{\left[\t_{43}^2(1+\t_1^2)+2x\t_{43}\t_{31}(1+\t_1\t_4)
+x^2 \t_{13}^2(1+\t_4^2)\right]}\,,
\non &&
 \label{a4}
\end{eqnarray}
where $\t_4$ is not necessarily $=\infty$, so the ordering of the points 
$\t_1$, $\t_3$ and $\t_4$ preserves cycling symmetry.

The integrand in \eq{a4} differs from the projective-covariant one 
only by the ugly factor 
\be
\frac{1}{(1+\t_1^2)(1+\t_3^2)} \frac{(\t_{43}+x \t_{31})^2} 
{\left[\t_{43}^2(1+\t_1^2)+2x\t_{43}\t_{31}(1+\t_1\t_4)
+x^2 \t_{13}^2(1+\t_4^2)\right]}
\ee
which we shall now see is not important because the integral over $\t_1$ and
$\t_3$ is divergent.

In order to regularize the integral, we proceed like in \eq{Gphib} changing
\be
|\t_i-\t_j|\to |\t_i-\t_j|+1-b\,, \qquad b\to1
\ee
and rescale $\t_i \to (1-b) \tilde \t_i$. The linearly divergent part of
the integral over $\t_1$ and $\t_3$ in \eq{a4} decouples from the integral
over $x$ and reads
\bea
&&\frac1{1-b}\left[
\int\limits_{-\infty}^{\tilde \t_4} d \tilde \t_3
\int\limits_{-\infty}^{\tilde \t_3} d \tilde \t_1
+\int\limits_{-\infty}^{\tilde \t_4} d \tilde \t_3
\int\limits^{+\infty}_{\tilde \t_4} d \tilde \t_1
+\int\limits^{+\infty}_{\tilde \t_4} d \tilde \t_3
\int\limits_{\tilde \t_4}^{\tilde \t_3} d \tilde \t_1\right]  
\nonumber \\* &&~~\times 
\frac1{(|\tilde \t_{43}|+1)(|\tilde \t_{31}|+1)(|\tilde \t_{41}|+1)} 
+{\cal O}\left((1-b)^0\right).
\label{a4div}
\eea
This integral is convergent and does not depend on $\tilde \t_4$ because of the
invariance under translations.

We have thus reproduced \eq{veneziano}, modulo a constant given by \eq{a4div},
with the straight Regge trajectory $\alpha(t)=\alpha' t$.
 The Regge behavior comes
about in the same way as when there is  the $PSL(2;\mathbb{R})$ symmetry.

%%\eop

\end{document}